\documentclass{pnastwo}
\usepackage{bm,helvet,times}
\usepackage[]{graphicx,color,epstopdf}
\usepackage{textcomp,colortbl}
\usepackage{amssymb,amsfonts,amsmath}
\usepackage[round,numbers,sort&compress]{natbib}
\usepackage{hyperref}
\usepackage{jabbrv}
\usepackage{color}
\usepackage{chngcntr}
\usepackage[T1, OT1]{fontenc} 
\DeclareTextSymbolDefault{\dh}{T1}  
\graphicspath{{figures/}}

\DefineJournalAbbreviation{Nature}{Nature}
\DefineJournalAbbreviation{Science}{Science}
\DefineJournalAbbreviation{Biosystems}{Biosystems}
\DefineJournalAbbreviation{Development}{Development}
\DefineJournalAbbreviation{Mycologia}{Mycologia}

\URL{www.pnas.org/cgi/doi/10.1073/pnas.1518527113}
\copyrightyear{2016}
\issuedate{Author Version}
\volume{113}
\issuenumber{20}

\begin{document}

\title{Coordinated Beating of Algal Flagella\\ is Mediated by Basal Coupling}

\author{Kirsty Y. Wan\affil{1}{Department of Applied Mathematics and Theoretical
Physics, University of Cambridge, Wilberforce Road, Cambridge CB3 0WA, UK} \and Raymond E. Goldstein\affil{1}{}}

\maketitle

\begin{article}

\begin{abstract}
Cilia and flagella often exhibit synchronized behavior; this includes phase-locking, as seen in {\it Chlamydomonas}, and metachronal wave formation in the 
respiratory cilia of higher organisms.
Since the observations by Gray and Rothschild of phase synchrony of nearby swimming spermatozoa, it has been 
a working hypothesis that synchrony arises from hydrodynamic interactions between beating filaments. 
Recent work on the dynamics of physically separated pairs of flagella isolated from the multicellular alga 
{\it Volvox} has shown that hydrodynamic coupling alone is sufficient to produce synchrony. 
However, the situation is more complex in unicellular organisms bearing few flagella. 
We show that flagella of \textit{Chlamydomonas} mutants deficient in filamentary connections between basal bodies 
display markedly different synchronization from the wildtype.
We perform micromanipulation on configurations of flagella and conclude that a mechanism, 
internal to the cell, must provide an additional flagellar coupling.
In naturally-occurring species with $4$, $8$ or even $16$ flagella, we find diverse symmetries
of basal-body positioning and of the flagellar apparatus that are coincident with specific gaits of flagellar actuation, 
suggesting that it is a competition between intracellular coupling and hydrodynamic interactions that ultimately 
determines the precise form of flagellar coordination in unicellular algae.

\end{abstract}


\section{Significance Statement} {\it In areas as diverse as developmental biology, physiology and
biomimetics there is great interest in understanding the mechanisms by
which active hair-like cellular appendages known as flagella or cilia
are brought into coordinated motion. The prevailing theoretical
hypothesis over many years is that fluid flows driven by beating
flagella provide the coupling that leads to synchronization, but this
is surprisingly inconsistent with certain experimentally observed
phenomena. Here we demonstrate the insufficiency of hydrodynamic
coupling in an evolutionarily significant range of unicellular algal
species bearing multiple flagella, and suggest the key additional
ingredient for precise coordination of flagellar beating is provided
by contractile fibers of the basal apparatus.}

\section{Introduction}
\dropcap{P}ossession of multiple cilia and flagella bestows significant evolutionary advantage upon living organisms 
only if these organelles can achieve coordination.
This may be for purposes of swimming \cite{Machemer1972,Goldstein2015}, feeding \cite{Orme2001}, or fluid 
transport \cite{Kramer-Zucker2005,Guirao2007}.
Multiciliation may have evolved first in single-celled microorganisms due to the propensity for hydrodynamic 
interactions to couple their motions, but was retained in higher organisms, occurring in such places as 
the murine brain \cite{Lechtreck2009} or human airway epithelia \cite{Smith2008}.
Since Sir James Gray first noted that ``automatic units'' of flagella beat in ``an orderly sequence'' when placed 
side by side \cite{Gray1928}, others have observed the tendency for nearby sperm cells to undulate in unison or aggregate 
\cite{Rothschild1949,Riedel2005}, and subsequently the possible hydrodynamic origins of this 
phenomenon have been the subject of extensive theoretical analyses \cite{Gueron1999, Guirao2007,Goldstein2015}. 
Despite this, the exclusiveness and universality of hydrodynamic effects in the coordination of neighboring cilia and flagella 
remains unclear.

We begin by considering one context in which hydrodynamic interactions are sufficient for 
synchrony \cite{Brumley2014}. The alga \textit{Volvox carteri} (VC) is perhaps the 
smallest colonial organism to exhibit cellular division of labor \cite{Solari2011}.
Adult spheroids possess two cell types: large germ cells interior of an extracellular matrix grow to form new 
colonies, while smaller somatic cells form a dense surface covering of flagella protruding into the medium, 
enabling swimming. 
These flagella generate waves of propulsion which despite lack of centralized or neuronal control (``coxless'') 
are coherent over the span of the organism \cite{Brumley2012}.
In addition, somatic cells isolated from their embedding colonies (Fig. \ref{fig:1}A) 
beat their flagella in synchrony when held sufficiently close to each other \cite{Brumley2014}.
Pairwise configurations of these flagella tend to synchronize in-phase (IP) when oriented with power strokes in the same direction, 
but antiphase (AP) when oriented in opposite directions, as predicted \cite{Leptos2013} if their mutual interaction were  
hydrodynamic. 
Yet, not all flagellar coordination observed in unicellular organisms can be explained thus. 
The lineage to which \textit{Volvox} belongs includes the common ancestor of the alga \textit{Chlamydomonas reinhardtii} (CR) (Fig.~\ref{fig:1}B), 
which swims with a familiar in-phase breaststroke with twin flagella that are developmentally positioned to 
beat in \textit{opposite} directions (Fig.~\ref{fig:1}C,D).
Yet, a \textit{Chlamydomonas} mutant with dysfunctional phototaxis switches stochastically the
actuation of its flagella between IP and AP modes \cite{Leptos2013, Wan2014}.
These observations led us to conjecture \cite{Leptos2013} that a mechanism, internal to the cell, must function to overcome 
hydrodynamic effects. 

Pairs of interacting flagella evoke no image more potent than Huygens' clocks \cite{Huygens1673}: two oscillating 
pendula may tend towards synchrony (or anti-synchrony) if attached to a common support, whose flexibility
providing the necessary coupling.
Here we present a diverse body of evidence for existence of a biophysical equivalent to this mechanical coupling, 
which in CR and related algae we propose is provided ultrastructurally by 
prominent fibers connecting pairs of basal bodies (BB) \cite{Ringo1967} that are known to have contractile properties. 
Such filamentary connections are absent in configurations of two pipette-held uniflagellate 
cells and defective in a class of CR mutants known as \textit{vfl} (Fig. \ref{fig:1}B).
We show in both cases that the synchronization states are markedly different from the wildtype breaststroke.

Seeking evidence for the generality of putative internal control of flagellar coupling in algal unicells, 
we use light microscopy, high-speed imaging and image-processing to elucidate the remarkable coordination 
strategies adopted by quadri-, octo-, and hexadecaflagellates, 
which possess networks of basal, interflagellar linkages that increase in complexity with flagella number. 
The flagellar apparatus, comprising BBs, connecting fibers, microtubular rootlets and the transition 
regions of axonemes, is among the most biochemically and morphologically complex structures occurring in eukaryotic 
flagellates \cite{InouyeBook}.  The significance of basal coupling relative to hydrodynamics is highlighted, especially 
in maintaining relative synchrony in diametrically-opposed pairs of flagella.
Our study reconciles species-specific swimming gaits across distinct genera of green algae with the geometry of flagellar placement and 
symmetries of their differing basal architecture -- so often a key phylogenetic character \cite{Systematics1984}.

While many features of eukaryotic flagellar axonemes are conserved from algae to mammals 
(e.g. composition by microtubules, motive force generation by dyneins, signal transduction by radial-spokes/central 
pair \cite{Wan2014}), far greater diversity exists in the coordination of multiple flagella.
Such strategies are vital not only in microswimmers bearing few flagella, 
but also in ciliary arrays. 
In mice defects in structures known as basal feet can cause ciliopathies \cite{Kunimoto2012}, 
while the striated (kinetodesmal) fibers in \textit{Tetrahymena} help maintain BB orientation and 
resist hydrodynamic stresses \cite{Galati2014}.
Insights from primitive flagellates may thus have significant broader implications.

\section{Results}
\subsection{Synchronization of Chlamydomonas flagella\label{sec:r1}}

\begin{figure}[t]
\centering
\includegraphics[width=0.46\textwidth]{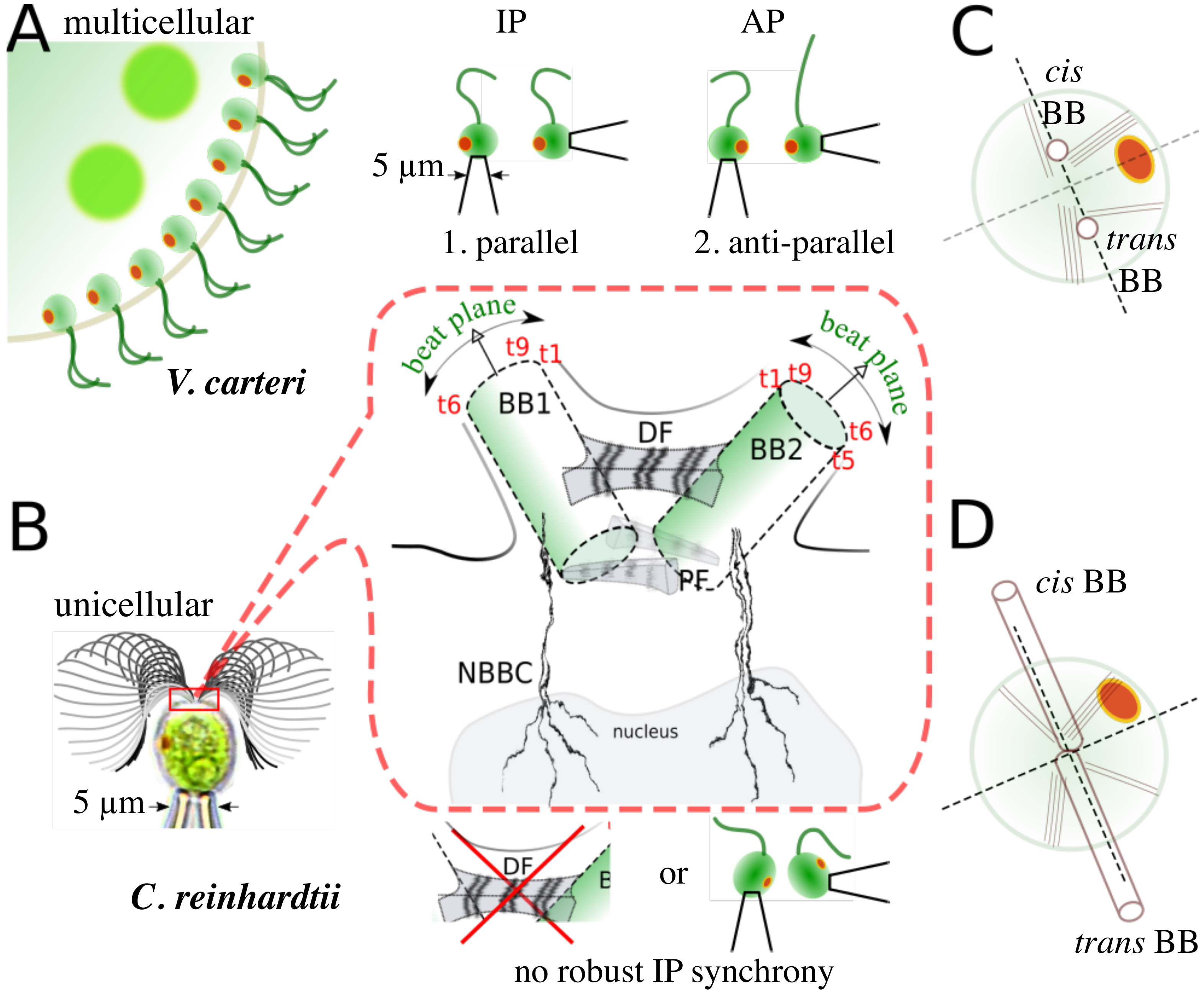}	
\caption{\small Flagellar synchronization in multi- vs uni- cellular algae. 
A) Pairs of isolated, somatic flagella of \textit{V. carteri} (VC) tend to synchronize either in in-phase (IP) or anti-phase (AP) 
depending on their relative orientation. 
B) \textit{C. reinhardtii} (CR) flagella maintain position $2$ yet swims a robust IP breaststroke that is lost
i) by mutation of the distal fiber (DF), and ii) in pairs of nearby uni-flagellate cells.
C+D) Ultrastructure comprising basal bodies (BBs), rootlets, and eyespot in VC and CR.
[PF - proximal fiber(s); NBBC - nuclear-basal-body connectors; t$1-9$: numbered microtubule triplets.]
\label{fig:1}}
\end{figure}

The basic configuration of two flagella appears in multiple lineages by convergent evolution, e.g. in 
the naked green alga \textit{Spermatozopsis} \cite{Mcfadden1987}, in gametes of the seaweed 
\textit{Ulva} \cite{Carl2014}, and in swarm cells of \textit{Myxomycetes} \cite{Gilbert1927}.
CR exemplifies the isokont condition. 
Cells ovoid, $\sim 5\,\mu$m radius, have flagella $\sim1.5\times$ body-length and distinguishable by BB age.
During cell division each daughter retains one BB from the mother \cite{Dieckmann2003} which becomes associated with the 
\textit{trans}-flagellum, while a second is assembled localizing near the eyespot and associates with the \textit{cis}-flagellum 
(Fig. \ref{fig:1}B,D).
When both flagella prescribe identical beats a nearly-planar breaststroke results, which is highly recurrent and stable to 
perturbations \cite{Wan2014b}.
Yet despite extensive research \cite{Ruffer1987,Goldstein2009,Bruot2012,Geyer2013,Quaranta2015} 
exactly how this IP breatstroke is achieved has remained elusive; 
coupling of the flagella pair may be by i) hydrodynamics, ii) drag-based feedback due to cell-body rocking, or iii) intracellular means. 

CR cells turn by modulation of bilateral symmetry. 
During phototaxis \cite{Witman1993} photons incident on the eyespot activate voltage-gated calcium channels 
which alter levels of intracellular calcium, leading to 
differential flagellar responses.
Ionic fluctuations (e.g. Ca$^{2+}$) alter not only the flagellar beat, 
but also the synchrony of a pair.
Gait changes involving transient loss of synchrony (called 'slips'),
 occur stochastically at rates sensitive to such environmental factors \cite{Lewin1952,Leptos2013,Wan2014b} as 
temperature, light, chemicals, hydrodynamics, and age of cell culture.
In free-swimming cells, slips can alter the balance of hydrodynamic drag on the cell body, 
producing a rocking motion that promotes subsequent resynchrony of flagella \cite{Geyer2013}, 
but this does not explain the robust IP synchrony in cells held immobilized on micropipettes \cite{Goldstein2009,Wan2014}, nor the 
motility of isolated and reactivated flagellar apparatuses \cite{Hyams1978}.
The altered beat during slips is analogous to the freestyle gait (AP in Fig.~1) characterized in the phototaxis 
mutant \textit{ptx1}, which stochastically transitions between IP and AP gaits \cite{Leptos2013,Wan2014}.
The dependence of CR flagellar synchronization state on physiology through temperature or ionic content of the medium \cite{Wan2014b} 
leads us now to the possibility for intracellular coupling of flagella. 

Early work \cite{Ringo1967} identified thick fibers connecting the two \textit{Chlamydomonas} BBs, including a 
$300\times250\times75$ nm$^3$ bilaterally symmetric distal fiber (DF), bearing complex striation 
patterns with a periodicity of $\sim\!80$ nm (Fig. \ref{fig:1}B).
Striation periodicity varies across species, and is changeable by chemical stimuli -- indicating active 
contractility \cite{Wright1983}.
The DF contains centrin, also found in NBBCs (Fig.~\ref{fig:1}B) which are involved in localization of BBs during cell division \cite{Wright1985}.
The two BBs have an identical structure of $9$ triplet microtubules which form a cartwheel arrangement \cite{Geimer2004}.
Importantly, the DF lies in the plane of flagellar beating, and furthermore attaches to each BB at the \textit{same} site 
relative to the beating direction of the corresponding flagellum (Fig.~\ref{fig:1}B). 
This inherent rotational symmetry makes the DF uniquely suited to coordinating the in-phase \textit{Chlamydomonas} breaststroke.

Hypothesizing a key role for the DF in CR flagellar synchrony we assess the motility of the
mutant \textit{vfl}$3$ (CC$1686$, \textit{Chlamydomonas} Center), with DFs missing, 
misaligned or incomplete in a large fraction of cells \cite{Wright1983}. 
Swimming is impaired -- many cells rotate in place at the chamber bottom.
In \textit{vfl}$3$ the number of flagella ($0-5$), their orientation and localization on the cell body, as well 
as cell size are abnormal.
BBs still occur in pairs,  but not every BB will nucleate a flagellum \cite{Hoops1984}, thus allowing flagella number to be odd.  
However no structural or behavioral defects were observed in the flagella \cite{Wright1983}.

\begin{figure*}[t]
\centering
\includegraphics[width=0.9\textwidth]{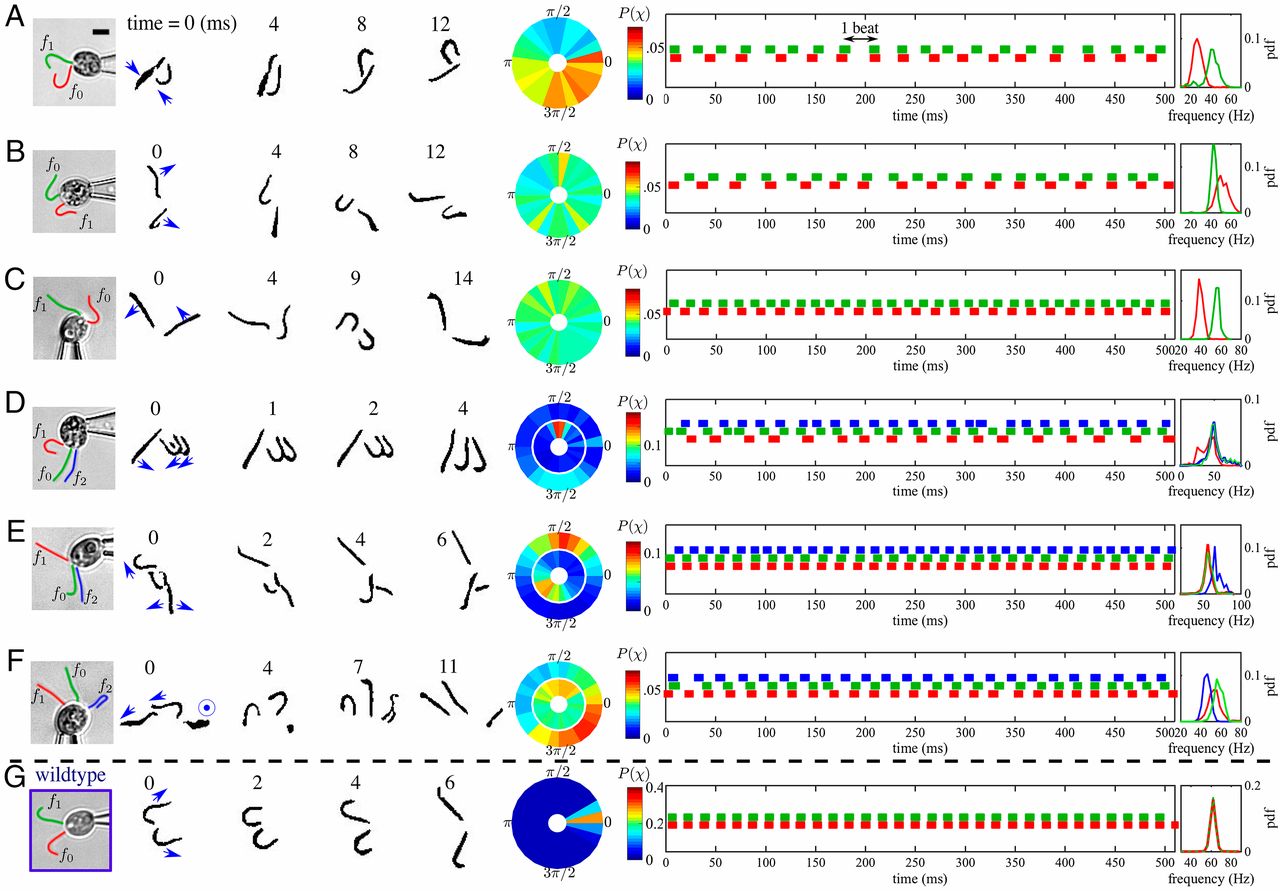}	
\caption{\small The CR mutant \textit{vfl3} has defective DF, abnormal flagella number and orientation. Scale bar: $5$ $\mu$m.
Shown are groups of $2$ or $3$ flagella in orientations of interest. (A,B) are toward or away-facing and flagella-like i.e. anti-parallel; 
(C) is cilia-like i.e. parallel; (D,E) exhibits clear hydrodynamic phase-locking of closely-separated parallel or anti-parallel pairs of flagella. 
Only (F) has a non-planar aspect: flagellum $f_2$ points out of the page.
Power stroke directions are indicated by the arrows.
Phase distributions ${\cal P}_0(\chi)$ of flagellum $f_{1,(2)}$ conditional on the phase of flagellum $f_0$ 
are shown on circular plots (in D-F: $f_1$ for the inner ring and $f_2$ for the outer).
For A-C, $S_{1,0}=0.20, 0.04, 0.04$, and for D-F, $(S_{1,0},S_{2,0}) = (0.30, 0.49), (0.53, 0.40), (0.02, 0.01)$.
In contrast for the wildtype, $S_{1,0}=0.96$ (panel G).
Discretized phases are plotted as "footprints" with length proportional to beat-cycle duration, with pdfs of beat frequencies.
\label{fig:2}}
\end{figure*}

A number of representative configurations of flagella occur in \textit{vfl3}, for 
cells bearing $2$ or $3$ flagella (Fig.~\ref{fig:2}A-F).
Fig.~\ref{fig:2}G presents the wildtype case.
We consider pairwise interactions between flagella.
For each flagellum, we extract a phase $\phi(t)$ from the high-speed imaging data by interpolating peaks in the standard 
deviation of pixel intensities measured across pre-defined regions of interest. 
\textit{vfl}$3$ flagellar beating frequencies are found to be more variable than the wildtype, so we elect to
determine phase synchrony between pairs of flagella via a stroboscopic approach. 
Given phases $\phi_1$, $\phi_2$ we wish to characterize the distribution ${\cal P}_C(\chi) $ of
$\chi = \phi_2\,\text{mod}\,2\pi |_{t: \phi_1\text{mod}\,2\pi =C}\,.$
Thus, the phase of flagellum $2$ is measured \textit{conditional} on the phase of flagellum $1$ attaining the value $C$.
From long timeseries we determine $\chi$ by binning $[0,2\pi]$ into $25$ equi-phase intervals centered 
around $\left\{C_k, k = 1,\cdots, 25\right\}$ to obtain the $N_k$ corresponding time points for which $\phi_1\text{mod}\,2\pi$ 
falls into the $k$th interval.
The distribution of this conditional phase $\chi^k:=\left\{\phi_2(t_i), i = 1,\cdots,N_k\right\}$, which is peaked when 
oscillators phase-lock and uniform when unsynchronized, can then be displayed on a circular plot by conversion 
to a colormap.
In Fig.~\ref{fig:2}, we take $k=1$. 
Phase vectors can be summed and averaged to define a synchronization index
\begin{equation}
	{\cal S} = \frac{1}{25}\sum_{k}^{25}\bigg|\frac{1}{N_k}\sum_{j}^{N_k}\exp(i\chi^k_j)\bigg|\,, \label{eqn:syncidx}
\end{equation}
where ${\cal S}_k=1$ (perfect synchrony) and ${\cal S}_k=0$ (no synchronization).

In \textit{vfl}$3$, steric interactions between nearby flagella (e.g. Fig. \ref{fig:2}A) 
can lead to intermittent beating and reduction in beat frequency. 
Even when measured beat frequencies differ for flagella on the same cell, periods of phase-locking are observed, 
which we attribute to hydrodynamic interactions \cite{Niedermayer2008,Leptos2013,Brumley2014} (see also SI Video $1$). 
In the tri-flagellates of Figs. \ref{fig:2}D\&E, beating of a given flagella pair becomes strongly coupled, 
with IP or respectively AP synchrony being preferred when flagella are oriented with power strokes parallel or respectively anti-parallel 
(compare $f_0$,$f_2$ in D with $f_0$,$f_2$ in E).
Even biflagellate \textit{vfl}$3$ cells with a native CR-like configuration cannot perform IP breaststrokes (e.g. Fig~\ref{fig:2}B).
In contrast wildtype CR flagella operating over a large frequency range are able achieve robust synchrony, 
despite intrinsic \textit{cis}/\textit{trans} frequency differences of up to $30\%$ during slips, 
or conditions of physiological stress such as deflagellation \cite{Wan2014}. 
Thus possession of functional or complete DFs appears necessary for CR flagellar synchrony.

\subsection{Micromanipulation of flagellate algae\label{sec:r2}}

Next we ask whether the CR breaststroke can instead be produced by hydrodynamic interactions.
Use of flagella belonging to \textit{different} cells offers a tractable alternative to removal by mutation of such physical connectors as the DF.
We construct, by micromanipulation, configurations of two flagella that cannot be coupled other than through the immersing fluid.

\begin{figure}[t]
\centering
\includegraphics[width=0.5\textwidth]{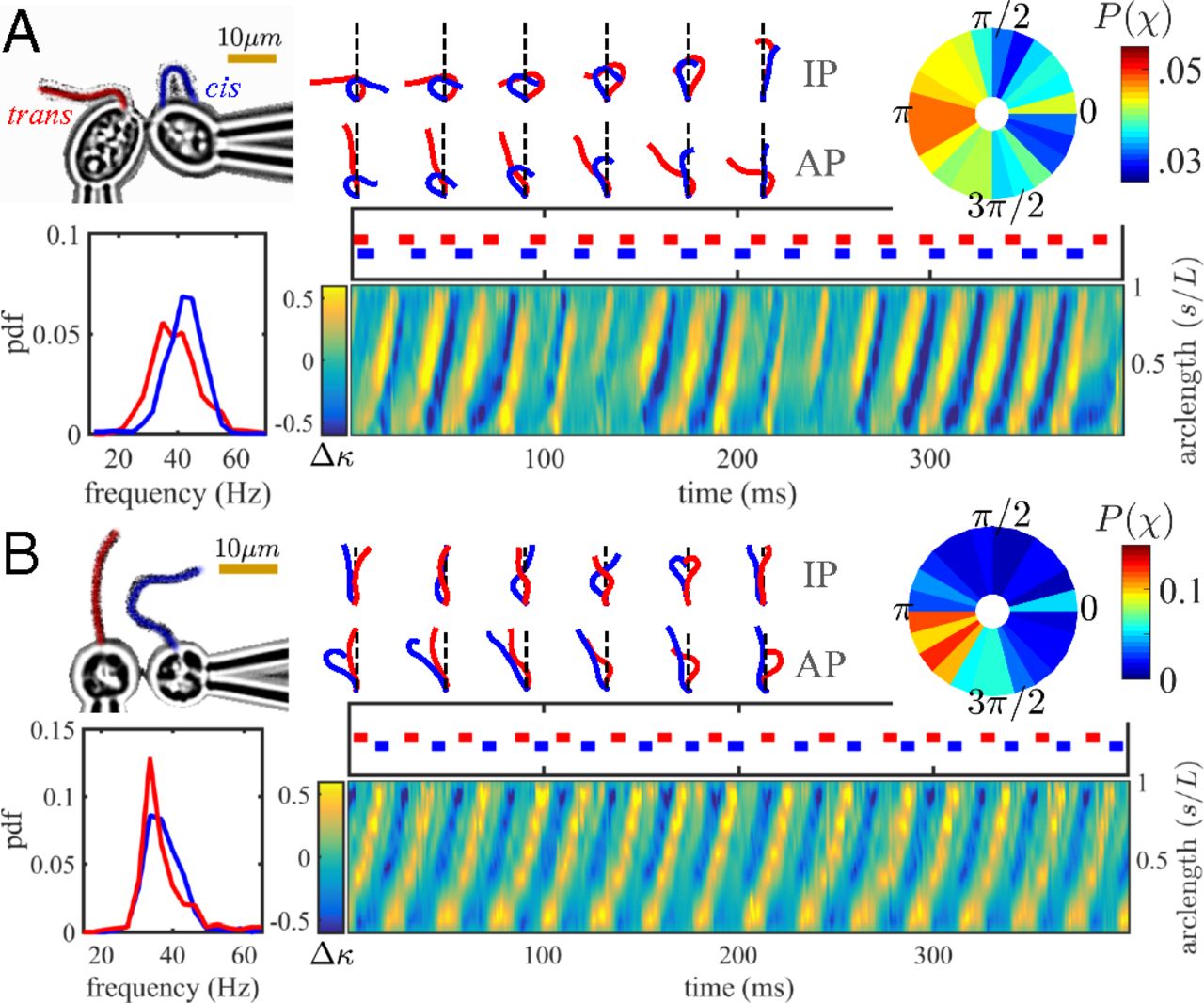}
\caption{\small Coupling A) two CR flagella (one \textit{cis}, one \textit{trans}) 
and B) two uni-flagellated VC somatic cells, both in anti-parallel configuration.
In both cases IP and AP states are observed, the AP being preferred. 
In B), hydrodynamic coupling is notably stronger. 
The pairwise curvature difference $\Delta\kappa$ ($\mu$m$^{-1}$) plotted on the same time axis as phase "footprints", 
shows propagating high-curvature bends which are coincident during IP, or alternating during AP.
\label{fig:3}}	 
\end{figure}

In Fig. \ref{fig:3}A, one flagellum was removed from each of two wildtype CR cells by careful mechanical shearing (Materials and Methods),
so that a CR-like arrangement comprising one \textit{cis} and one \textit{trans} flagellum is assembled.
Despite similarity with the wildtype configuration, no sustained IP breaststrokes were obtained.
Closely separated ($<\!\!5$ $\mu$m) pairs exhibit periods of phase locking. 
Beat frequencies of these flagella are found to be more noisy than their counterparts in intact CR cells, 
and consequently measured phase-locking is not robust ($S=0.08$). 
The conditional (stroboscopic) phase $\chi$ (Fig. \ref{fig:3}A) is peaked weakly about $\pi$ indicating a tendency for AP synchronization, 
but the IP state is also possible.
This bistability is not species-specific; 
we rendered uniflagellate [see also \cite{Brumley2014}] pairs of pipette-held VC somatic cells (normally biflagellate)
and placed them in a similar configuration (Fig. \ref{fig:3}B).
Analysis of the resulting pairwise flagellar interactions indicates a strong preference for AP synchronization,
though the IP is again observed (see SI Videos~$2$ \& $3$). 
Accordingly, the phase stroboscope is now strongly peaked near $\chi=\pi$ (with $S = 0.67$).

The existence, stability, and frequency of IP and AP states are wholly consistent
with a basic theory \cite{Niedermayer2008} which models a pair of hydrodynamically coupled flagella as 
beads rotating on springs with compliant radii $R$ separated by distance $\ell$ (see SI Text). 
Assuming $R\ll \ell$, either IP or AP states of synchrony are predicted to be stable 
depending on whether the beads are co- or respectively counter-rotating \cite{Niedermayer2008,Leptos2013}.
Thus, CR-like configurations should tend to AP synchrony.
However in our experiments, flagella often come into such close proximity during certain phases of their beat cycles that 
the far-field assumptions of the original model breakdown.
Non-local hydrodynamic interactions between different portions of flagella must now be considered for
the true flagellar geometry (rather than in a phase-reduced bead model). 
In particular undulating filaments can be driven by fluid-structure interactions into either IP or AP oscillating modes 
depending on initial relative phase \cite{Elfring2009}.
The inherent stochasticity of flagellar beating \cite{Wan2014b} thus leads to transitions between IP and AP states (Fig. \ref{fig:3}).
Hydrodynamic effects are notably stronger in the case of VC than CR, due to reduced screening by a smaller cell body and a distinctive 
upward tilt of the flagellar beat envelope. 
During evolution to multicellularity, this latter adjustment of BB orientation (Figs. \ref{fig:1}C\&D) facilitates beating of flagella confined within a spherical colony.
The CR mutant \textit{ptx1} also displays noisy transitions between IP and AP gaits \cite{Leptos2013} due to an unknown mutation, 
the implications of this we shall return to later (see Discussion).

The similarity of two flagellar waveforms in any given state (IP or AP) can be compared.
For ease of visualization, waveforms discretized at equidistant points 
are ordered from base to tip: $\mathbf{f}^\text{L}_i$, $\mathbf{f}^\text{R}_i$ for $i=1,\cdots N_p$, and rescaled to uniform total length.
These are rotated by ${\cal T}(\alpha)$ through angle $\alpha$ between the horizontal and line of offset between the BBs. 
We denote by
$\mathbf{f}^\text{L}\rightarrow\mathbf{f'}^\text{L}={\cal T}(\mathbf{f}^\text{L}-\mathbf{m})$ and
$\mathbf{f}^\text{R}\rightarrow\mathbf{f'}^\text{R}={\cal T}(\mathbf{f}^\text{R}-\mathbf{m})$, with 
$\mathbf{m}=(\mathbf{f}^\text{L}_1+\mathbf{f}^\text{R}_1)/2$, 
and compare the resulting shape symmetries during IP and AP states (Figs. \ref{fig:3}A\&B, stacked).
The synchronization index of Eq. \ref{eqn:syncidx}, while suitable for identifying presence or absence of phase synchrony, 
does not discriminate between AP or IP states. 
Instead, we compute the pairwise curvature difference $\Delta\kappa(t,s) = \kappa_L-\kappa_R$ as a function of time ($t$) 
and normalized arclength ($s$), where $\kappa_L$, $\kappa_R$ are \textit{signed} curvatures for the left and right flagellum 
according to their respective power stroke directions. 
During IP and AP synchrony, $\Delta\kappa$ exhibits a wave pattern at the common or phase-locked frequency.
Principal or high-curvature bends propagate from flagellum base to tip in accordance with the mechanism of flagellar beating in these species 
($0.897\pm0.213$ mm$/$s for CR, and $0.914\pm0.207$ mm$/$s for VC).      
In summary, we have established in two species that robust IP synchrony akin to the CR-breaststroke 
does not arise from hydrodynamic coupling between two co-planarly beating flagella arranged in a CR-like configuration, even when beat frequencies are comparable. 

Moreover the propensity for algal flagella to be deformable by hydrodynamic loading \cite{Brumley2014} -- e.g. flows generated by nearby flagella, suggests an internal coupling must be present to compensate in such cases where fluid interactions are contrary to the desired mode of propulsion by flagella in the organism. 
Given this delicate interplay, will a dramatic perturbation to the state of hydrodynamic interactions between flagella affect their \textit{native} mode of coordination? 
For this we require an organism with more than two flagella. 
\textit{Tetraselmis} is a thecate quadriflagellate, and amenable to micromanipulation.
Fig.~\ref{fig:5}A depicts a pipette-immobilized \textit{Tetraselmis} cell with flagella free to beat in a pattern 
qualitatively similar to free-swimming cells observed under identical conditions (see SI Video $4$), 
in which flagella maintain a transverse gallop (next section, and Fig.~\ref{fig:4}D). 
One flagellum was then trapped inside a second pipette with suction so as to completely stall its beating, with minimal disruption to the cell.
Flagellar dynamics were monitored and interflagellar correlation functions computed (Fig.~\ref{fig:4}B), 
showing that the prior beat patterns continue (Fig.~\ref{fig:4}C). 
The small increase in beat frequency in the remaining flagella ($\sim5\%$) is consistent with calcium-induced frequency elevation by mechanosensation. 
This remarkable ability for the cell to sustain its coordination pattern strongly implicates internal beat modulation.

\subsection{Symmetries of the algal flagellar apparatus}
Such a hypothesis brings us now to further detailed study of a large number of lesser-known species that have differing or more complex basal architectures 
and which in turn, we find to display varied and novel flagellar coordination strategies.
While it is believed that Volvocine green algae (including VC) derived from \textit{Chlamydomomonas}-like ancestors, 
the general classification of Viridiplanta (green plants) has undergone repeated revisions 
due to the enormous variability that exists in the form and structure exhibited by its member species. 
Features, both developmental (mitosis, cytokinesis) and morphological (number, structure, arrangement of flagella, nature of body coverings such as 
scales and theca), have served as key diagnostics for mapping the likely phylogenetic relationships existing between species 
\cite{Stewart1978, Melkonian1980}.

We selected unicellular species of evolutionary interest to exemplify configurations of $2,4,8$ or even $16$ flagella (SI Videos $6,7\&8$).
These include (Fig. \ref{fig:5}) distinct genera of bi- and quadri- flagellates -- both occurring abundantly in nature, 
the rare octoflagellate marine Prasinophyte known as \textit{Pyramimonas octopus}, and its relative \textit{Pyramimonas cyrtoptera} -- the only species known with $16$ flagella \cite{Daugbjerg1992}.
Indeed only three \textit{Pyramimonas} species have $8$ flagella during all or parts of its cell cycle \cite{Conrad1939, Hargraves1980, Hori1987}. 
Several species belong to the Prasinophytes -- a polyphyletic class united through lack of similarity with either 
of the main clades (Chlorophyta and Streptophyta), whose  $>\!16$ genera and $>\!160$ species \cite{Sym1991a} display a remarkable variability
in flagella number and arrangement that is ideal for the present study.

\begin{figure}[b]
	\centering
	\includegraphics[width=0.42\textwidth]{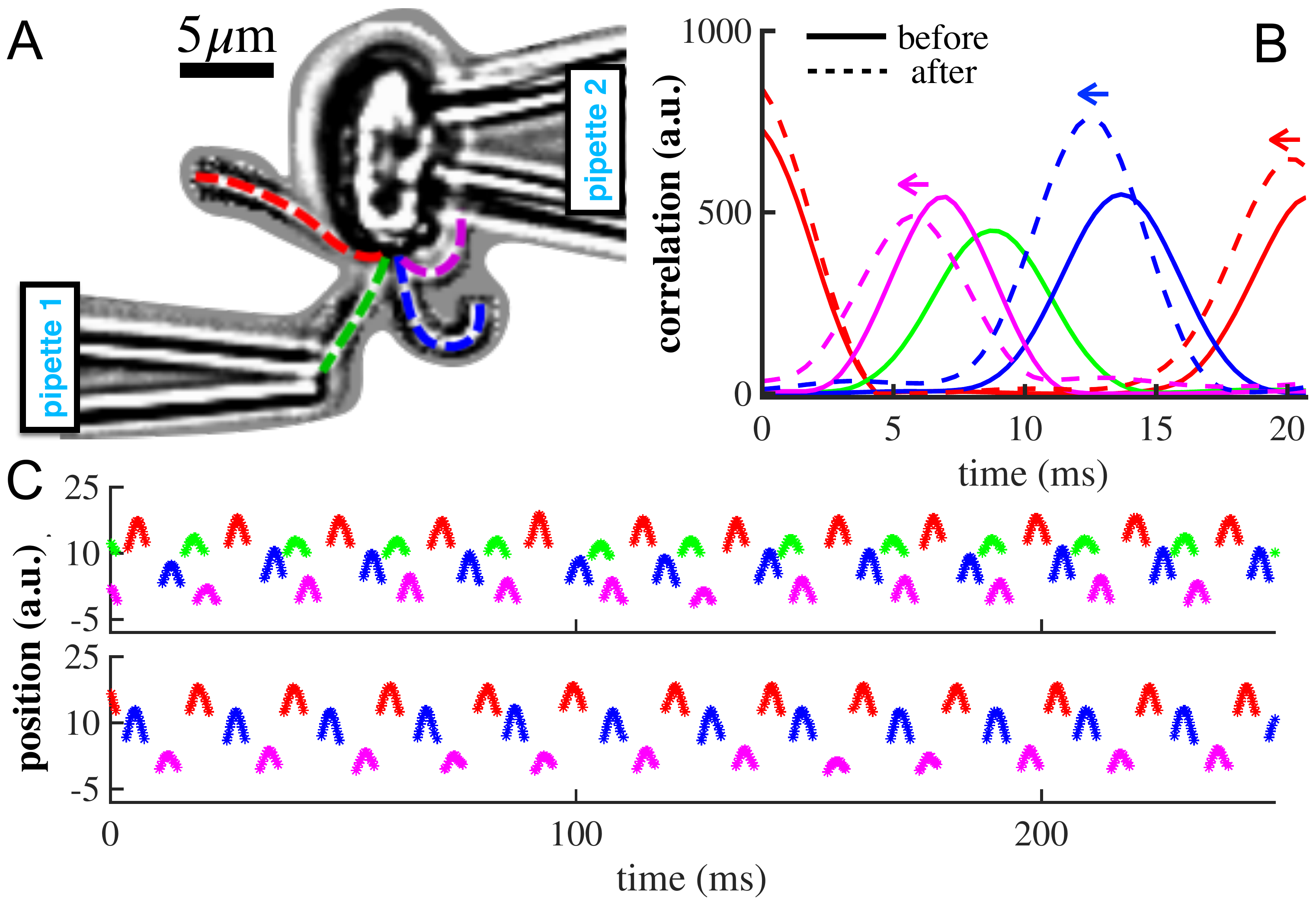}
	\caption{\small A) Stalling the flagellum beat in {\it Tetraselmis suecica}.
		B) Beat correlations for each flagellum relative to a reference flagellum (red) computed before and after manipulation, 
		shows period shift but no change in the order of flagella actuation. 
		C) Timeseries or "footprints" delineate positions of flagella (labelled as in A) before (above) and after (below) manipulation. 
		\label{fig:4}}
\end{figure}  	

Distinct quadriflagellate gaits were identified, involving particular phase relations between flagella.
An analogy may be drawn with quadruped locomotion (Fig. \ref{fig:5}\&\ref{fig:6}).
For measured flagellar phases $\psi_j$ (flagellum index $j$) we compute the 
matrix $\Delta_{ij} = \psi_i-\psi_j \quad(i,j = 1,\dots,4)$, where
$\Delta_{ij} = \Delta_{ji}$, $\Delta_{ii} = 0$, and $\Delta_{ik} = \Delta_{ij}+\Delta_{jk}$.
Each gait is then associated with a 3-tuple of phase differences: $[\Delta_{12}\,\Delta_{13}\,\Delta_{14}]$.
For instance, \textit{P. parkeae} swims with two pairs of precisely alternating breaststrokes akin to the `trot' of a horse (Fig. \ref{fig:6}A, SI Video $6$),
its $4$ isokont flagella insert anteriorly into an apical pit, emerging in a cruciate arrangement typical of quadriflagellates (Fig.~\ref{fig:6}, type I).
The phase relation $[\Delta_{12}\, \Delta_{13}\,\Delta_{14}] = [\pi\,0\,\pi]$ is seen in both free-swimming and micropipette-held cells.
A Chlorophyte alga \textit{Polytomella} sp. \textit{parva} (Fig. \ref{fig:5}), was also found to display this gait. 
Two further gaits (SI Video $6$) occur in the type {\it Pyramimonas} species \textit{P. tetrarhynchus} \cite{Manton1968}, a 
freshwater alga. 
The first we term the `pronk', where all flagella synchronize with zero phase difference (Fig.~\ref{fig:6}B).
In the second, flagella beat in a sequence typical of the transverse gallop in quadrupeds (Fig. \ref{fig:5}).
\textit{P. tetrarhynchus} swims preferentially with the latter gait, while pronking can occur when the cell navigates near walls/obstacles, or when changing direction. 
The rotary gallop, with flagella beating CCW in orderly sequence (Fig.~\ref{fig:6}C, SI Video $6$) occurs in the Volvocale \textit{Carteria crucifera}.
Finally in \textit{Tetraselmis} (recall Fig.~\ref{fig:4}) the flagella separate distally into pairs (Fig.~\ref{fig:6}, type II). 
Cells display the transverse gallop when free-swimming or pipette-held despite strong hydrodynamic interactions within each pair. 
An alternate synchronous gait of four flagella has been reported in this species \cite{Salisbury1978}, but was not observed under our experimental conditions. 

\begin{figure*}[t]
	\centering
	\includegraphics[width=0.83\textwidth]{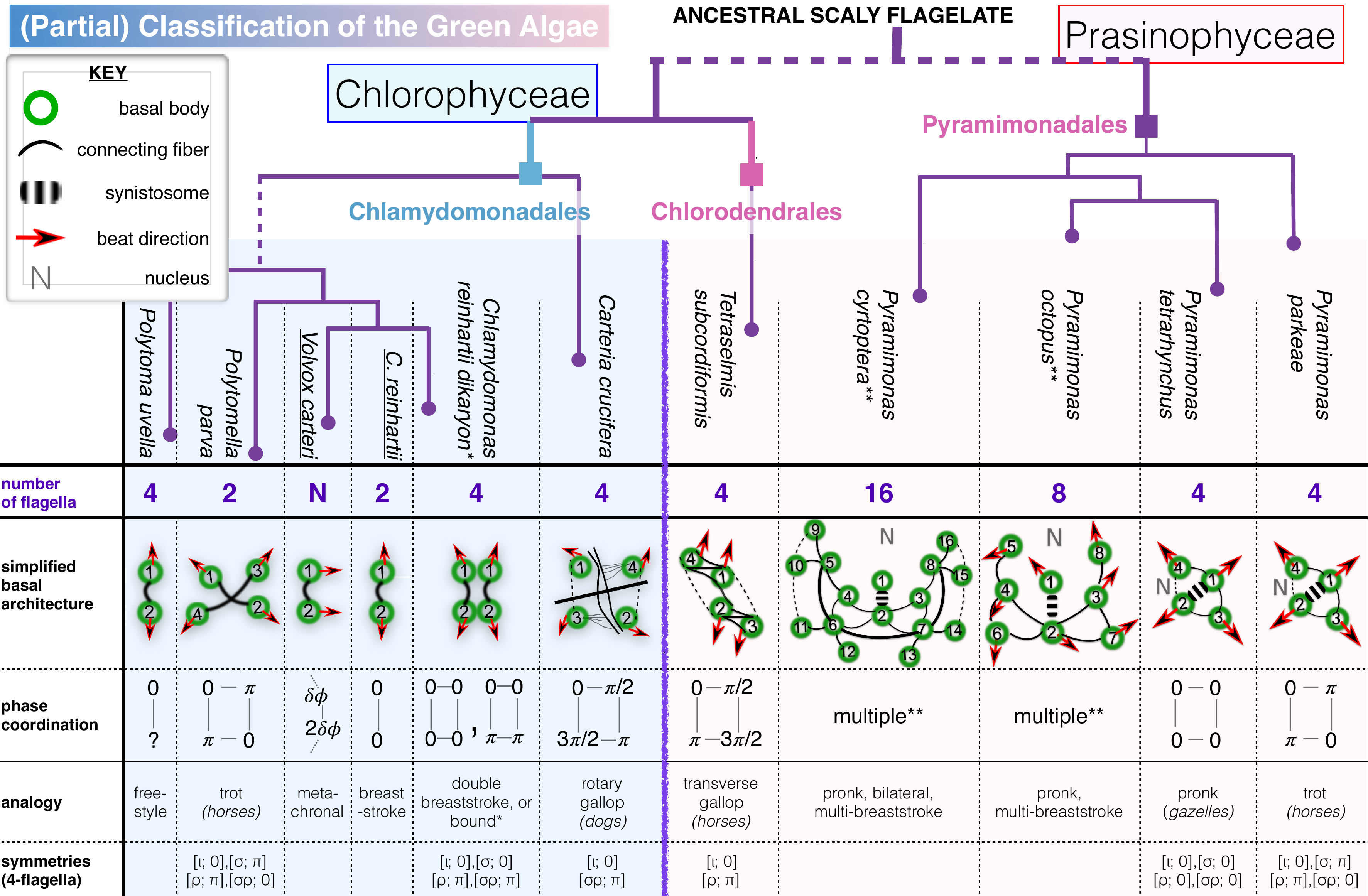}
	\caption{\small 
		Algal species are compared in terms of phylogeny, number and orientation of flagella, 
		arrangement of BBs/basal architecture (see also SI Text), 
		and patterns of coordination defined by the relative phases between flagella.
		Vertical lines approximate the relative phylogenetic distance from a putative flagellate ancestor (SI Text) (only partial branchings are shown). 
	    Free-swimming quadriflagellate gaits are readily identified with quadruped locomotion, revealing the symmetries of an underlying oscillator network. 
		[* Quadriflagellate dikaryons of CR gametes perform a double breaststroke gait with the same symmetries as the `pronk' (SI Video 5), 
		but this gives way to a `bound' gait when both sets of flagella undergo phase slips simultaneously.
		** See SI Videos $7$\&$8$.]
		\label{fig:5}}
\end{figure*}  
\begin{figure*}[t]
	\centering
	\includegraphics[width=0.86\textwidth]{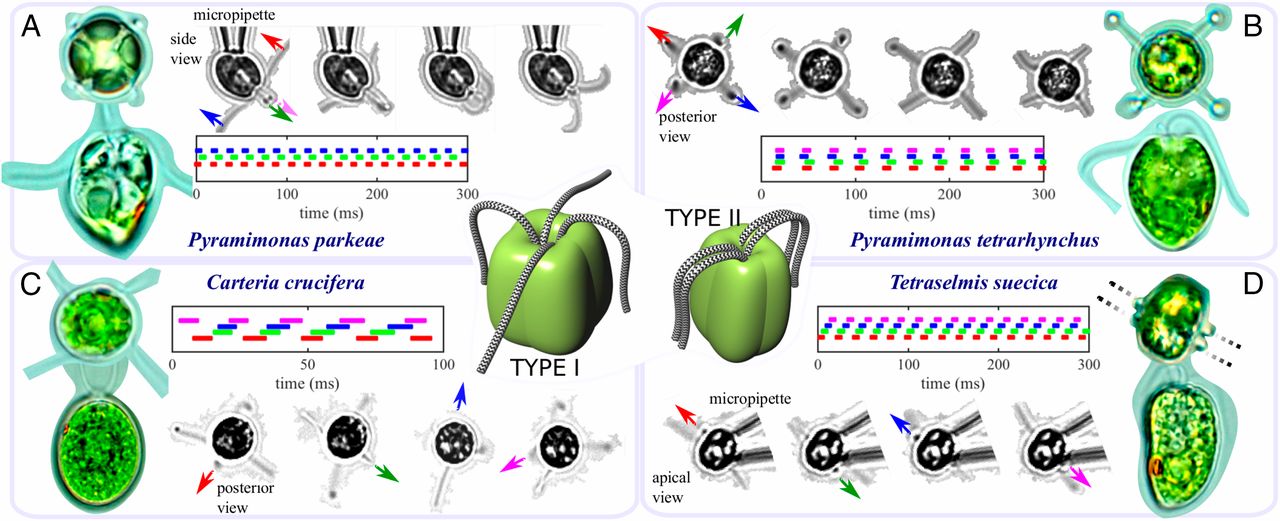}
	\caption{\small 
		In free-living quadriflagellates, flagella can be arranged in one of two possible configurations (types I,II) -- type II is unusual.
		Gaits of coordination are diverse and species-specific, including the trot (A), pronk (B), rotary (C) and transverse (D) gallops.
		Representative species imaged from top and side, show locations of eyespots and chloroplast structures. 
		Timeseries of phases are measured for pipette-held cells in A,D, and free-swimming cells in B,C. 
		\label{fig:6}}
\end{figure*} 

Ordinarily, motive gaits whereby the limbs of a quadruped or arthropod are actuated in precise patterns are
produced by networks of coupled oscillators controlled by a central pattern generator (CPG) or equivalent \cite{Collins1993, Schoner1990}.
Analogously, can we relate symmetries of the flagellar apparatus to gait symmetries, 
with contractile filaments providing the putative coupling?
We focus on quadriflagellates, where an abundance of species makes possible comparative study (Fig. \ref{fig:6}). 
Spatial symmetries of flagella $\mathbf{X}_j$ (indexed by $j=1,2,3,4$) 
are represented as permutations $\sigma$ of $\left\{1,2,3,4\right\}$, so that $\mathbf{X}_k(t)=\mathbf{X}_{\sigma(k)}(t)$ 
for all times $t$.
Periodic gaits also possess temporal symmetries: if $T$ is the gait period, then for the $k$th oscillator 
$\mathbf{X}_k(t)=\mathbf{X}_k(t+T)$ for all $t$. 
We normalize $T$ to $2\pi$ and consider invariance of flagella under phase shifts $\phi$ taken modulo $T/2\pi$; 
the pair $[\sigma, \phi]$ denotes a spatial-temporal symmetry of the $\{\mathbf{X}_j\}$, where 
\begin{equation}
\mathbf{X}_k(t) = \mathbf{X}_{\sigma(k)}(t-\phi) ~.
\end{equation}
and $k=1,\cdots ,4$.
The set of spatiotemporal symmetries may admit a (symmetry) group under composition:
\begin{equation}
[\sigma_1; \phi_1] \circ [\sigma_2; \phi_2] = [\sigma_1\sigma_2; \phi_1+\phi_2] ~,
\end{equation}
to be matched with known quadruped/quadriflagellate gaits. 
For a basal morphology of a rectangularly symmetric network \cite{Collins1993} with distinct lengthwise and crosswise couplings 
-- in Fig.~\ref{fig:5} (phase coordination) represented by lines of differing lengths, 
spatial symmetries include $\iota$ (the identity, fix everything), $\sigma = (12)(34)$ (reflect in $y$-axis), 
$\rho = (13)(24)$ (reflect in $x$-axis), and $\sigma\rho = (14)(23)$ (interchange of diagonals).
In Fig.~\ref{fig:5}, we attach named gaits and associated symmetry groups to the species in which they are observed.
Additionally some quadriflagellates display a `stand' gait (a transient rest phase where no flagellum beats); this 
has the largest number of symmetries: $[\iota; \phi], [\sigma; \phi],[\rho; \phi],[\sigma\rho; \phi]$, for arbitrary $\phi$.

Can such a network resemble coupling of algal flagella?
Despite significant variation across species, linkages or roots connecting BBs are key systematic characters. 
These can be microtubular or fibrillar. 
Microtubular roots, which were probably asymmetric in very early flagellates \cite{Systematics1984}, 
position BBs and attached flagella during development (two per BB: termed left, right). 
The right root is generally $2$-stranded, and together with the left-root ($X$-stranded) 
form an X-2-X-2 cruciate system characteristic of advanced green algae [$X=4$ in CR \cite{Ringo1967}].
Only one absolute configuration of BBs exists for each species and its mirror-symmetric form is not possible \cite{Okelly1983a}. 
For instance in CR two BBs emerge at $70-90^\circ$ with a clockwise (CW) offset characteristic of advanced biflagellate 
algae (Fig. \ref{fig:1}D), in contrast many evolutionarily more primitive flagellates have BBs oriented with a counter-clockwise (CCW)
offset \cite{Systematics1984}. 
Fibrillar roots, classified as system I or II \cite{Melkonian1980} become more numerous with flagella number.
These are generally contractile, likely contributing to interflagellar coupling.
Each BB is unique up to the imbrication of its member tubules -- a constant positional relationship pertains between its two 
roots and a principal connecting fiber linking the two ontogenetically oldest BBs, labelled 1,2 in keeping with convention \cite{Okelly1983a,Moestrup1989}.
It is this fiber that is mutated in \textit{vfl}$3$ mutants (recall Fig. \ref{fig:2}).

Take a quadriflagellate species for which the basal architecture is known, and consider its associated swimming mode.
In \textit{P. parkeae} (Fig. \ref{fig:6}A), a prominent (striated) distal fiber called the synistosome links BBs $1,2$ only \cite{Norris1975}, 
so that the coupling is different between different pairs.
In the advanced heterotroph \textit{Polytomella parva} which swims with a comparable gait, 
flagella form opposing V-shaped pairs with different coupling between pairs \cite{Brown1976}. 
In \textit{Tetraselmis}, the flagella separate distally into two nearly collinear pairs 
with BBs forming a single zig-zag array (Fig.~\ref{fig:5}), in a state thought to have arisen from rotation of two of the flagellar roots in an ancestral quadriflagellate. 
Transfibers which are functionally related to the \textit{Chlamydomonad} PF and DF, connect alternate BBs,
while BBs within the same pair are linked by Z-shaped struts \cite{Salisbury1981}, 
emphasizing a diagonal connectivity which may explain its transverse gallop (Fig. \ref{fig:6}D).
In contrast the rotary gallop, prominent in \textit{C. crucifera} (Fig. \ref{fig:6}C) is more consistent with a 
square symmetric network involving near identical connectivity between neighboring flagella, than a rectangular one.
Indeed the flagellar apparatus in this species has been shown to exhibit unusual rotational symmetry: BBs insert into an anterior papilla at the corners of a square 
in a cruciate pattern (class II \textit{sensu} Lembi \cite{Lembi1975}) and are tilted unidirectionally in contrast to the conventional V-shapes found in \textit{Chlamydomonas} and \textit{Polytomella}. (Here DFs rather than link directly to BBs, attach to rigid electron-dense rods extending between them \cite{Lembi1975}.)
Finally, a different network of couplings appears when two biflagellate CR gametes fuse during sexual reproduction \cite{Harris2009} 
to form a transiently quadriflagellate dikaryon (similar to configuration II, Fig. \ref{fig:6}).
Here, original DFs between \textit{cis} and \textit{trans} remain but new fibers do not form between pairs of flagella of unlike mating type. 
Strong hydrodynamic coupling due to their physical proximity results in a striking \textit{double} bilateral breaststroke (SI Video $5$).
A `bound'-like gait can appear if both sets of CR flagella slip together (Fig. \ref{fig:5}).

\begin{figure}[b]
	\centering	
	\includegraphics[width=0.37\textwidth]{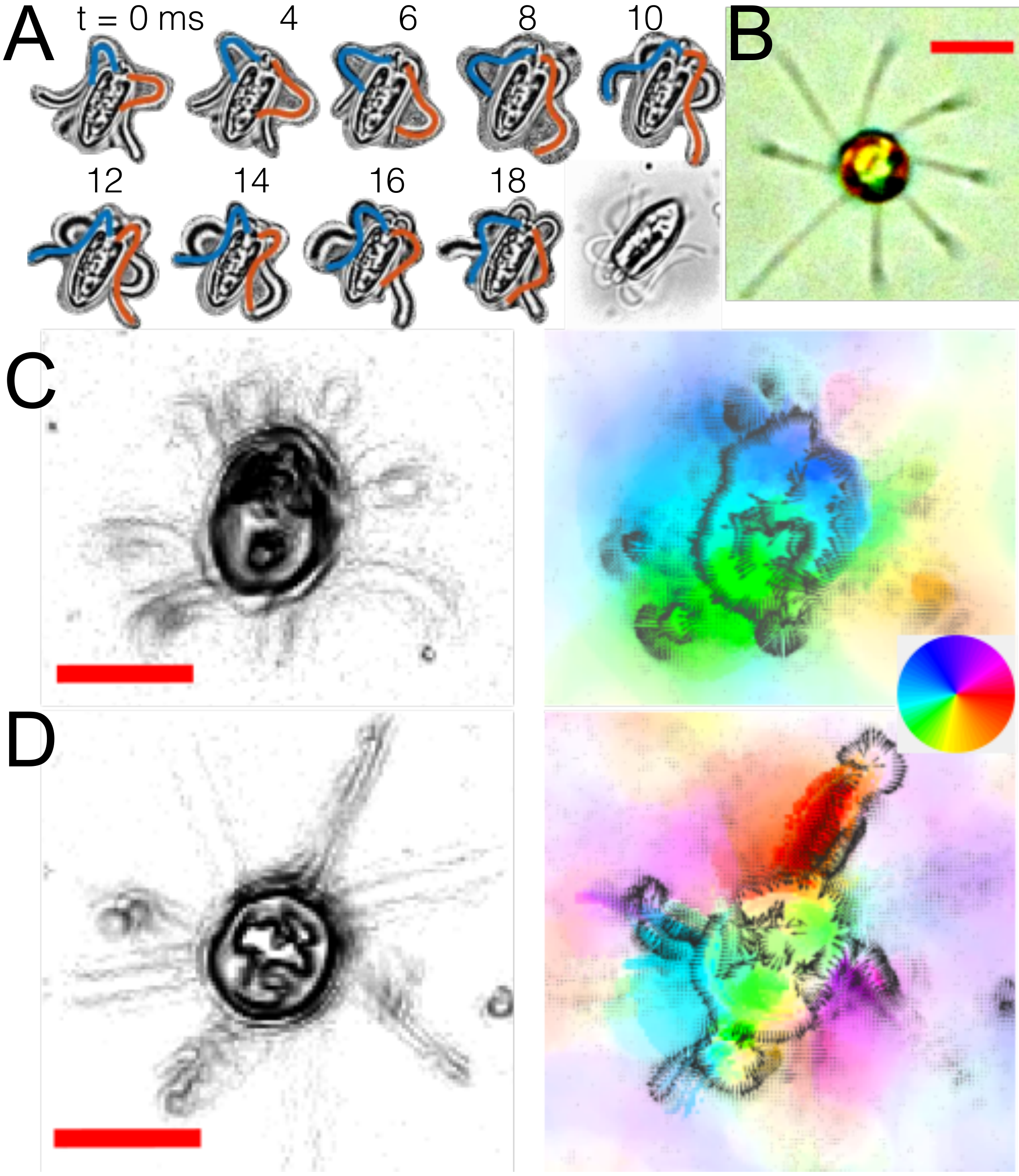}
	\caption{\small \textit{P. octopus} cell A) swims using multiple pairs of breaststrokes: 
		highlighted are waveforms of one synchronous pair.
		In B), cell is at rest in a `stand' gait in which no flagellum is active. 
		C-D) Gradient images (left) identify the active subset of flagella (all, or $4$ of $8$), 
		optical flow fields (right) show decay of beating-induced flow disturbance.
		Scale bars: $10$~$\mu$m.
		\label{fig:7}}
\end{figure} 

Thus, algal motility appears to be constrained by the form of underlying coupling provided by a species-specific configuration of BBs and connecting fibers.
The case of octo- and hexadeca- flagellates swimming is more complex (SI Video 7,8), 
limited by the number of species available for study, and presence of a large number of fibrous structures whose identities remain unknown. 
A consistent numbering system for BBs  has greatly facilitated the study of flagellar transformation 
in algae with many flagella, which we adopt \cite{Okelly1983a,Moestrup1989}.
As in other \textit{Pyramimonas} BBs $1,2$ remain connected by a large synistosome (Fig. \ref{fig:5}).
In \textit{P. octopus} up to $60$ individual fibers establish specific connections with specific BB triplets, 
in addition to $6$-$8$ rhizoplasts linking the basal apparatus to the nucleus.
Numerous rhizoplasts and connecting fibers also exists in \textit{P. cyrtoptera} but were never resolved 
fully because of the untimely death of T. Hori \cite{ProtistologyBook}.
In \textit{P. octopus} (Fig. \ref{fig:7}) BBs arrange in a diamond (partially open to allow nuclear migration during mitosis). 
BB duplication is semi-conservative \cite{Beech1991}: $8$ new BBs form peripheral to the existing ones during cell division \cite{Moestrup1989}. 
The innermost BBs ($1,2$) assume the new position $1$ in the two daughters (i.e. full maturation) after round 
$1$ of cell division, but BBs $3,4$ and $5$-$8$ only reach maturation respectively after rounds 
$2$, $3$ \cite{Hori1987,Moestrup1989}. All $2^n$ BBs in a given cell reaches and thereafter remains at position 
$1$ by the $n$th generation (cf \textit{P. cyrtoptera} \cite{Daugbjerg1992}). 

\begin{figure*}
	\centering
	\includegraphics[width=0.98\textwidth]{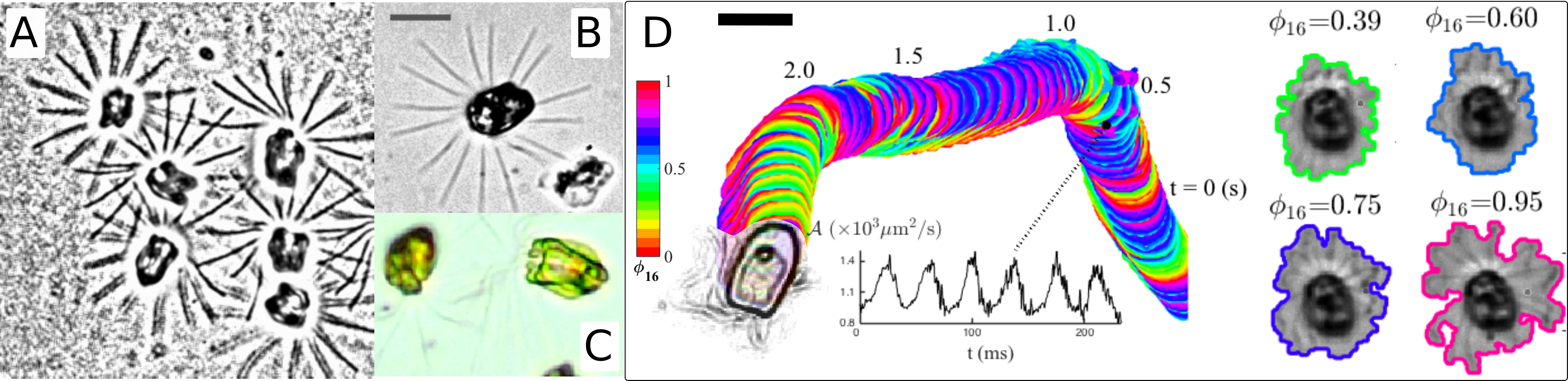}
	\caption{\small In the rare Arctic species \textit{P. cyrtoptera} flagella can remain at rest for seconds in a `stand' gait (A-B).
		Cells appear yellow-green and lobed (C).
		(D) Pronking involving all $16$ flagella (SI Video 8) is a common swimming gait. 
		The cell body is tracked along a typical trajectory and colored by normalized phase $\phi_{16}$ (colorbar), computed from 
		the area ${\cal A}(t)$ bounding the flagella which expands and contracts according to the periodicity of flagellar beating (shown here at $4$ representative phases).
		Inset: time series of ${\cal A}$ over $6$ successive cycles. 
		Scale bars: $20$~$\mu$m.
		\label{fig:8}}
\end{figure*}

In electron micrographs of \textit{P. octopus}, flagellar beating is oriented in CCW sequence \cite{Hori1987,Moestrup1989} 
consistent with observed CW body rotation (viewed from above). 
Cells $\approx 20$ $\mu$m in length, with yet longer flagella whose distal portions fold and recurve by the cell,
swim at $200\sim300$ $\mu$m$/$s along helical paths.
Diametrically opposite groups (usually but not exclusively pairs) of flagella beat synchronously with ordered phase shifts between groups (Fig.~\ref{fig:7}A, SI Video 7).  
In-phase coupling appears more robust in distinguished pairs (SI Video 7)
reflecting prominent symmetries of the basal architecture: 
e.g. a large synistosome links BBs $1,2$ whose flagella beat in opposite directions (Fig.~\ref{fig:5}).
Compatible with its benthic nature, \textit{P. octopus} display the `stand' gait introduced previously (Fig. \ref{fig:7}B) .
Swimming motion resumes spontaneously when all $8$ flagella reactivate (pronking), and the normal gait is reestablished within only $2$ or 
$3$ beats. 
A transient state in which beating occurs in $4$ of $8$ flagella (staggered) is observed Fig. \ref{fig:7}D: 
in which actively beating flagella are identified by summing successive gradient images (in contrast ih Fig. \ref{fig:7}C all $8$ flagella are in motion).
We visualise the flow disturbance imparted by flagellar beating using optical flow to monitor the migration of image pixels between frames, 
and estimate flow directions and magnitudes \cite{Sun2014} (also Computer Vision toolbox, MATLAB, The MathWorks Inc.).
Components of the $2D$ flow field were determined by minimizing an objective function subject to certain global (smoothness) 
or local (fidelity) constraints, and pixels delineating the same object are likely to have constant intensity. 
Unlike conventional PIV methods, this removes the need to seed the medium with foreign particles or tracers, 
and fluid flow due to motion can be obtained directly from optical data. 
Flow strengths decay rapidly away from boundaries of flagella as well as the cell body which is also in motion (Figs.~\ref{fig:7}C\&D, right).
More generally, gait transitions in these primitive algae can be dramatic and involve unusual flagellar coordination in response to external stimuli, 
and shall be explored further in a separate manuscript. 
Here, we reserve special attention to the \textit{P. cyrtoptera} pronking gait, which displays striking hydrodynamic interactions.

\subsection{Hydrodynamic synchronization in a hexadecaflagellate}
\textit{P. cyrtoptera} is the only hexadecaflagellate known \cite{Daugbjerg1994}. 
It is an Arctic species thought to have evolved when \textit{P. octopus} failed to divide after duplication of chloroplasts and flagella.
Measuring up to $40$ $\mu$m, this is the largest \textit{Pyramimonas} ever recorded.
Its $16$ flagella ($32$ when dividing), which are longer than the cell and emerge radially from a deep anterior flagellar pit (Figs. \ref{fig:8}A\&B),
are used by the organism to attach to icy surfaces.
The name \textit{P. cyrtoptera} is derived from the Greek for \textit{cyrtos} meaning ``curved'' and \textit{pteron} 
meaning ``wing'', in reference to cell morphology.
Light microscopy reveals a lobed structure with two pairs of split eyespots and the presence of two chloroplasts (Fig. \ref{fig:8}C).
\textit{P. cyrtoptera} cells are stenothermal and euryhaline: growth becomes limited above 7-8$^\circ$C.
Once removed from their natural habitat where temperature variations are typically $<\!2^\circ$C, cell cultures often 
prove fragile and difficult to maintain in the lab. 

Hexadecaflagellate swimming presents an intriguing circumstance in which the distance of separation between 
flagella is so small that hydrodynamic coupling is inevitable. 
When fully splayed (Fig. \ref{fig:8}B) the $16$ flagella are separated on average by $360/16 = 22.5^\circ$ 
or $7.85$ $\mu$m measured at a radial distance of $20$ $\mu$m from the cell, compared to $360/8=45^\circ$ and twice this 
distance in \textit{P. octopus}.
In \textit{P. cyrtoptera} the interflagellar distance is thus far below the critical length to achieve synchronization by hydrodynamics \cite{Brumley2014}.
Strong hydrodynamic interactions are evident between some or even all flagella (SI Video 8), none more so than during the pronk.
Instead of tracking individual flagella (which beat in synchrony), we use the area ${\cal A}(t)$ bounding the flagella as proxy for phase of beat cycle (normalized):
\begin{equation}
\phi_{16}(t)= \left(\frac{t-t_n}{t_{n+1}-t_n}\right) \quad\quad [t_{n}\leq t\leq t_{n+1}]~,
\end{equation}
where $t_n$ are discrete marker events corresponding to local maxima of ${\cal A}(t)$.
Pronking occurs at a $~\sim25$ Hz (Fig. \ref{fig:8}D).
Significant hydrodynamic effects are also evidenced in cells swimming against surfaces, 
where flagella exhibit symplectic metachronal waves (SI Text).

\section{Discussion}
\subsection{Insufficiency of hydrodynamics} 
The phenomenon of cooperative or synchronous beating of cilia and flagella has received growing attention, 
with hydrodynamic interactions historically assumed to be the major source of coupling.
It is only recently \cite{Leptos2013,Geyer2013,Quaranta2015} that researchers have begun to question this long-standing belief. 
Our approach was motivated in part by the freestyle gait (AP) in the CR phototaxis mutant \textit{ptx1}, 
realizing that the wildtype IP breaststroke cannot be reconciled with hydrodynamic theory \cite{Leptos2013, Wan2014}.
An additional ingredient, internal to the cell, must be maintaining IP synchrony in CR. 
The finding that entrainment of CR flagella by periodic external flows only occurs at frequencies close to the natural 
beat frequency and strengths greatly in excess of physiological values led Quaranta \textit{et al} to a similar conclusion \cite{Quaranta2015}.
The DF likely couples CR flagella, providing a degree of freedom that can reorient a flagellum at the BB through its contraction.
In isolated and reactivated flagella apparatuses for example, the DF constricts in response to elevated extracellular calcium 
to reduce the opening angle between the two BBs \cite{Hyams1978}.
Since BBs nucleate/anchor flagella \textit{and} function as centrioles during cell division, 
the DF can also be affected by mutations in BB duplication and segregation \cite{Wright1983}.
This brings us back to the unusual flagellar coordination in \textit{ptx1}, 
which is thought to possess two \textit{trans}-like flagella \cite{Okita2005}.
If BB signalling or connectivity is disrupted or weakened in \textit{ptx1}, stochastic IP/AP transitions can result when
hydrodynamic interactions compete with a reduced intracellular coupling \cite{Leptos2013}. 
Indeed, for all their DF defects, flagella of \textit{vfl3} do not synchronize or only hydrodynamically when very close together (Fig. \ref{fig:2}).
Future work should seek to examine flagellar apparatuses of \textit{ptx1} under electron microscopy.
The failure of hydrodynamics to synchronize two flagella in a CR-like configuration, let alone two functionally distinct flagella such as a \textit{cis} and a \textit{trans} \cite{Kamiya1984,Wan2014}, was shown by micromanipulation of two pipette-held cells (Fig. \ref{fig:2}).

The diversity of coordination gaits in flagellate species (Fig.~\ref{fig:5}) implicates internal coupling as a generic remedy for this insufficiency of hydrodynamics.
Indeed, non-uniqueness of stable quadriflagellate gaits for even identical configurations of $4$ flagella is incompatible with existence of a single hydrodynamic mode.
In some species a number of gait bifurcations can occur during free-swimming, involving modification or even cessation of beating in one or more flagella suggesting coordination is an active process. 
The significant perturbation to the hydrodynamic landscape resulting from immobilizing one flagellum in \textit{T. suecica} was found to have little effect on the native coordination of the remainder (Fig.~\ref{fig:4}).
Thus symmetries of flagellar gaits are much more species- than drag- dependent.

\subsection{Intracellular coupling of flagella by contractile fibers}
Gaits are defined by the relative phase between oscillators, which in the analogy of multi-legged locomotion, 
may be produced by CPG or pacemaker signals which in algae we conjecture to be mediated by the basal architecture.
Flagellar apparatuses imaged by electron microscopy reveal species-specific networks of connections 
which increase in complexity with flagella number \cite{InouyeBook}.
Symmetries of an underlying network of structural couplings \cite{Collins1993} likely translate downstream into symmetries of 
observed multiflagellate gaits (Fig. \ref{fig:5}).
The BB from which the flagellum nucleates is a center for conduction of morphogenetic and sensory information between 
flagella and other intracellular organelles. 
Although BBs are not essential for flagellar function (isolated axonemes continue to beat when reactivated in 
ATP \cite{Bessen1980,Mukundan2014}), the contractility of inter-BB connections may contribute to coordination \cite{Melkonian1983a}.
In CR robust NBBCs descending near the DF link BBs to the nucleus (Fig.~\ref{fig:1}B) remain intact even 
after detergent lysis treatment \cite{Wright1985}, and can be induced by calcium to undergo significant contraction \cite{Salisbury1988}.
Similarly rhizoplasts of scaly algae including {\it Pyramimonas} and {\it Tetraselmis} can contract and relax cyclically \cite{Melkonian1980,Salisbury1978}.
These species display frequent directional changes (mechanoshock) that may be mediated by the extraordinary 
contractility of rhizoplasts \cite{Salisbury1978}, with normal coordination after abrupt gait changes rapidly reestablished. 
Fibrillar structures under tension experience much distortion during \textit{active} beating, as observed from misalignment of 
fiber striation patterns \cite{Gibbons1960}. 
Contraction during live beating is ATP-dependent in {\it Polytomella} \cite{White1981}, occurs in real-time in paralyzed 
flagella and temperature sensitive mutants of \textit{Chlamydomonas} \cite{Hayashi1998}, and to an extraordinary degree in 
{\it Microthamnion} zoospores \cite{Watson1975}.
ATPase activity has also been identified in the rod cells of the human eye where a large striated root attaches to the 
BB of a short (non-motile) cilium \cite{Matsusaka1967}.
Attachment sites of contractile fibers also exhibit great specificity, in most species to specific microtubule triplets and disc complexes.
In \textit{P. octopus} contractile fibers attach to the "weaker" side of BBs (triplets 6-9), and may function to 
pull the flagellum back from each power stroke during its unique multi-breaststroke gait.

\subsection{Evolution of multiflagellation among Viridiplantae}
As an appendage, cilia, flagella and its $9+2$ axoneme prevails across eukaryotes and especially the green algae; 
yet beyond the universality of this basic machinery much variability (Fig. \ref{fig:5}) persists in the placement of organelles, 
form of flagellar insertion, and diversity of flagellar coordination. 
The basal apparatus is that rare structure which is both universally distributed and stable enough to infer homology 
across large phylogenetic distances, and yet variable enough to distinguish between different lineages. 
Representative species considered here express $2^n$ flagella, with much conservatism in the bi- or quadri- 
flagellate condition.
Since an early flagellate phagocytosed a prokaryote (the future chloroplast) $>$1 billion years ago,
green algae have evolved photosynthesis and autotrophism \cite{Cavaliersmith1982}. 
Their radiation and division into the Streptophyta and Chlorophyta \cite{Leliaert2012} has been corroborated by 
modern high-throughput chloroplast genome sequencing \cite{Buchheim1996}. 
Occupying a basal phylogenetic position are morphologically diverse species 
of freshwater and marine \textit{Prasinophytes}, including the \textit{Pyramimonas} species \cite{Systematics1984} studied here.
These have conspicuous body and flagella scales which are precursors of theca and \textit{Volvocalean} cell walls \cite{Okelly1983a}.
In particular \textit{P. tetrarhynchus}, \textit{P. octopus}, and \textit{P. cyrtoptera} are assigned according to morphological characters \cite{McFadden1986} 
to the same subgenus, in which accelerated rates of evolution were confirmed by cladistic analysis of \textit{rbcL} gene sequences (Fig.~\ref{fig:5}). 
Changes in basal ultrastructure were major events \cite{Systematics1984} with the quadriflagellate condition arising multiple times, 
in fact it is the quadriflagellates (e.g. \textit{Carteria}) and not Chlamydomonad-type biflagellates that are considered basal to advanced Volvocales \cite{Buchheim1996,Nozaki2003}.
The advanced heterotrophic alga \textit{Polytomella}, thought to have evolved by cell doubling along a direct line of descent from CR \cite{Smith2014}, 
displays the same trot gait as the ancestral \textit{P. parkeae}.
In these cases, convergent ultrastructural modifications coincident with multiplicity and doubling of BBs 
may have evolved to enable strong coupling between opposite flagella pairs.

The sparsity of species bearing $>4$ localized flagella (\textit{P. octopus}, \textit{P. cyrtoptera}) 
may stem from the difficulty and activity costs of a flagellar apparatus capable of maintaining coordination from within 
despite external effects such as hydrodynamics.
\textit{P. cyrtoptera} exemplifies an intermediate between few to many flagella 
($16$ is the highest number ever reported in a phytoflagellate \cite{Daugbjerg1992}), 
and able to exploit hydrodynamics for swimming in a novel manner (Fig. \ref{fig:5}).
The energetic gains of such cooperativity may have inspired derivation of multiflagellated colonial volvocales from unicellular ancestors. 
Opportunities for fluid-mediated flagellar coordination and metachronism imposes new constraints on the configuration of flagella and BBs. 
In VC somatic cells for example, BBs reorient during early stages of development to become parallel (Fig. \ref{fig:1}C), 
while biflagellate VC sperm cells (required to swim independently) retain the primitive V-formation.

\subsection{Implications for active control of flagellar coordination}
From the comparative studies carried out in this work we conclude that the physical principle for coordinating 
collective ciliary beating in \textit{Volvox} or \textit{Paramecium} differs from that responsible for defining 
precise patterns of beating in unicellular microswimmers bearing only few flagella. 
In the former case hydrodynamic coupling between flagella is sufficient \cite{Brumley2014}, 
but in the latter (especially for obligate autotrophs) there is far greater incentive for efficient swimming 
to be robust \textit{in spite of} hydrodynamic perturbations. 
Even in arrays of mammalian cilia, ciliary roots and basal feet structures continue to provide additional resistance to 
fluid stresses \cite{Brooks2014,Galati2014}.   

Rapid changes in cellular structure that are of fundamental evolutionary interest may have arisen
in the first instance in green flagellates than higher organisms. 
At the base of flagella in these algae are found diverse networks of interconnecting filaments that are not only responsible for 
anchorage and placement of flagella but which must now also be implicated in defining the symmetries of flagellar coordination.
Fiber contractility can produce elastic coupling between BBs to force nearby flagella into 
modes of synchrony (IP or AP) that oppose hydrodynamic influences \cite{CosentinoLagomarsino2003, Leptos2013}.
This elasticity may be actively modulated, highlighting a direct correlation between cellular physiology and 
flagellar beating that has already been identified \cite{Wan2014b,Ma2014}.
Further insights into such processes will certainly require additional modelling and experimentation.
The rapidity with which patterns of synchrony can change is suggestive of transduction by electrical signals or ionic 
currents \cite{Harz1991}, which may be effected from cell interior to flagella by changes in the state of contraction 
or relaxation of connecting fibers.
Striations of algal rhizoplasts are even biochemically mutable in a manner reminiscent of mammalian muscle. 
We are therefore led to suggest that a parallel evolution of neuromuscular control of appendages may have occurred much earlier than 
previously thought \cite{Salisbury1978,Jekely2015}.
                        
\section{Materials and Methods}

\subsection{Culturing and growth of algae} Below are brief descriptions of the protocols for species 
whose flagellar dynamics are studied here.

{\it Volvox.} \textit{V. carteri} was prepared as described elsewhere \cite{Brumley2014}.
The remaining species, unless otherwise specified, were maintained under controlled illumination 
on $14:10$ day/night cycles, and at a constant temperature of 22$^\circ$C (incubation chamber, Binder).

{\it Pyramimonas.}
Marine species obtained from the Scandanavian Culture Collection of Algae and Protozoa,
K-0006 \textit{P. parkeae} R.E. Norris et B.R. Pearson 1975 (subgenus \textit{Trichocystis}), 
K-0001 \textit{P. octopus} Moestrup et Aa. Kristiansen 1987 (subgenus \textit{Pyramimonas}), and K-0382 
\textit{P. cyrtoptera} Daugbjerg 1992 (subgenus \textit{Pyramimonas}), were cultured in TL30 medium \cite{SCCAP_media}. 
Of these, \textit{P. cyrtoptera} is an Arctic species and was cultured at $4^\circ$C. 
A fourth \textit{Pyramimonas}, K-0002 \textit{P. tetrarhynchus} Schmarda 1850 (type species) 
is a freshwater species, and was grown in an enriched soil medium NF2 \cite{SCCAP_media}. 

{\it Tetraselmis.}
Marine species \textit{T. suecica} (gift from University of Cambridge Department of Plant Sciences) and 
\textit{T. subcordiformis} (CCAP 116/1A), were cultured in the f/2 medium \cite{CCAP_media}.

{\it Polytoma.}
\textit{P. uvella} Ehrenberg 1832 (CCAP 62/2A) was grown in polytoma medium (comprising 2\% sodium acetate trihydrate, 
1\% yeast extract and 1\% bacterial tryptone \cite{CCAP_media}).

{\it Polytomella.}
Two species (CCAP 63/1 and CCAP 63/3) were maintained on a biphasic soil/water medium \cite{CCAP_media}.

{\it Carteria.}
\textit{C. crucifera} Korschikov ex Pascher (1927) from CCAP (8/7C) was grown in a modified Bold basal medium \cite{CCAP_media}.

{\it Chlamydomonas.}
\textit{C. reinhardtii} strains were obtained from the Chlamydomonas Collection, wildtype CC125, and variable flagella 
mutant {\it vfl3} (CC1686), and grown photoautotrophically in liquid culture (Tris-Acetate Phosphate).

{\it Production of quadriflagellate dikaryons.}
High-mating efficiency strains of \textit{C. reinhardtii} CC$620$ (mt$^+$), CC$621$ (mt$^-$) were obtained from the Chlamydomonas Collection, and grown photoautotrophically in nitrogen-free TAP to induce formation of motile gametic 
cells of both mating types. Fusing of gametes occurred under constant white light illumination. 
\subsection{Manipulation of viscosity}
To facilitate identification of flagella in certain species, the viscosity of the medium was increased by addition of Methyl cellulose (M7027, Sigma Aldrich, 15 cP) to slow down cell rotation and translation rates.
\subsection{Microscopy and micromanipulation}
The capture of single cells are as described elsewhere \cite{Wan2014,Wan2014b,Brumley2012,Brumley2014}.
For Fig.~\ref{fig:3}A caught CR cells were examined under the light microscope to identify the eyespot and thus \textit{cis} and \textit{trans} flagella;
the correct flagellum was then carefully removed using a second pipette with smaller inner diameter.

\begin{acknowledgments}
This work is supported by a Junior Research Fellowship from Magdalene College Cambridge (KYW) and a Wellcome Trust 
Senior Investigator Award (REG).
We thank Kyriacos Leptos and Marco Polin for discussions relating to the contractility of 
the basal apparatus at an early stage of this work, and the possible insights provided by the vfl class of mutants, 
Fran\c{c}ois Peaudecerf for kindly providing the CR gametes, and Stephanie H{\"o}hn for valuable comments on the manuscript. 
\end{acknowledgments}

\small
\bibliographystyle{pnas2011}
\bibliography{coupling}

\begin{thebibliography}{10}

\bibitem{Machemer1972}
Machemer H (1972) Ciliary activity and origin of metachromy in
  \textit{Paramecium} - effects of increased viscosity.
\newblock {\em\JournalTitle{Journal of Experimental Biology}} 57(1):239--259.

\bibitem{Goldstein2015}
Goldstein RE (2015) Green algae as model organisms for biological fluid
  dynamics.
\newblock {\em\JournalTitle{Annual Review of Fluid Mechanics}} 47:343--375.

\bibitem{Orme2001}
Orme BAA, Otto SR, Blake JR (2001) Enhanced efficiency of feeding and mixing
  due to chaotic flow patterns around choanoflagellates.
\newblock {\em\JournalTitle{IMA Journal of Mathematics Applied in Medicine and
  Biology}} 18(3):293--325.

\bibitem{Kramer-Zucker2005}
Kramer-Zucker AG et~al. (2005) Cilia-driven fluid flow in the zebrafish
  pronephros, brain and kupffer's vesicle is required for normal organogenesis.
\newblock {\em\JournalTitle{Development}} 132(8):1907--1921.

\bibitem{Guirao2007}
Guirao B, Joanny JF (2007) Spontaneous creation of macroscopic flow and
  metachronal waves in an array of cilia.
\newblock {\em\JournalTitle{Biophysical Journal}} 92(6):1900--1917.

\bibitem{Lechtreck2009}
Lechtreck KF, Sanderson MJ, Witman GB (2009) High-speed digital imaging of
  ependymal cilia in the murine brain.
\newblock {\em\JournalTitle{Cilia: Structure and Motility}} 91:255--264.

\bibitem{Smith2008}
Smith DJ, Gaffney EA, Blake JR (2008) Modelling mucociliary clearance.
\newblock {\em\JournalTitle{Respiratory Physiology \& Neurobiology}}
  163(1-3):178--188.

\bibitem{Gray1928}
Gray J (1928) {\em Ciliary Movement}.
\newblock (Cambridge University Press).

\bibitem{Rothschild1949}
Rothschild (1949) Measurement of sperm activity before artificial insemination.
\newblock {\em\JournalTitle{Nature}} 163(4140):358--359.

\bibitem{Riedel2005}
Riedel IH, Kruse K, Howard J (2005) A self-organized vortex array of
  hydrodynamically entrained sperm cells.
\newblock {\em\JournalTitle{Science}} 309(5732):300--303.

\bibitem{Gueron1999}
Gueron S, Levit-Gurevich K (1999) Energetic considerations of ciliary beating
  and the advantage of metachronal coordination.
\newblock {\em\JournalTitle{Proceedings of the National Academy of Sciences of
  the United States of America}} 96(22):12240--12245.

\bibitem{Brumley2014}
Brumley DR, Wan KY, Polin M, Goldstein RE (2014) Flagellar synchronization
  through direct hydrodynamic interactions.
\newblock {\em\JournalTitle{elife}} 3:e02750.

\bibitem{Solari2011}
Solari CA et~al. (2011) Flagellar phenotypic plasticity in volvocalean algae
  correlates with \text{P\'eclet} number.
\newblock {\em\JournalTitle{Journal of the Royal Society Interface}}
  8(63):1409--1417.

\bibitem{Brumley2012}
Brumley DR, Polin M, Pedley TJ, Goldstein RE (2012) Hydrodynamic
  synchronization and metachronal waves on the surface of the colonial alga
  \textit{Volvox carteri}.
\newblock {\em\JournalTitle{Phys. Rev. Lett.}} 109(26):268102.

\bibitem{Leptos2013}
Leptos KC et~al. (2013) Antiphase synchronization in a flagellar-dominance
  mutant of \textit{Chlamydomonas}.
\newblock {\em\JournalTitle{Phys. Rev. Lett.}} 111:158101.

\bibitem{Wan2014}
Wan KY, Leptos KC, Goldstein RE (2014) Lag, lock, sync, slip: the many "phases"
  of weakly coupled flagella.
\newblock {\em\JournalTitle{Journal of the Royal Society Interface}}
  11:20131160.

\bibitem{Huygens1673}
Huygens C (1673) {\em Horologium Oscillatorium, sive, De motu pendulorum ad
  horologia aptato demostrationes geometricae}.
\newblock (F. Muguet,).

\bibitem{Ringo1967}
Ringo DL (1967) Flagellar motion and fine structure of flagellar apparatus in
  \textit{Chlamydomonas}.
\newblock {\em\JournalTitle{Journal of Cell Biology}} 33(3):543--\&.

\bibitem{InouyeBook}
Inouye I (1993) Flagella and flagellar apparatuses of algae.
\newblock {\em\JournalTitle{Ultrastructure of microalgae}} pp. 99--133.

\bibitem{Systematics1984}
Irvine D, John D, eds. (1984) {\em The Systematics Association special volume
  Systematics of the green algae}, Systematics Association.
\newblock (Academic Press) Vol.{}~27.

\bibitem{Kunimoto2012}
Kunimoto K et~al. (2012) Coordinated ciliary beating requires odf2-mediated
  polarization of basal bodies via basal feet.
\newblock {\em\JournalTitle{Cell}} 148(1-2):189--200.

\bibitem{Galati2014}
Galati FD et~al. (2014) Disap-dependent striated fiber elongation is required
  to organize ciliary arrays.
\newblock {\em\JournalTitle{Journal of Cell Biology}} 207(6):705--715.

\bibitem{Mcfadden1987}
Mcfadden GI, Schulze D, Surek B, Salisbury JL, Melkonian M (1987) Basal body
  reorientation mediated by a \text{Ca}$^{2+}$-modulated contractile protein.
\newblock {\em\JournalTitle{Journal of Cell Biology}} 105(2):903--912.

\bibitem{Carl2014}
Carl C, de~Nys R, Lawton RJ, Paul NA (2014) Methods for the induction
  reproduction in a species of filamentous ulva.
\newblock {\em\JournalTitle{PLOS One}} 9(5):e97396.

\bibitem{Gilbert1927}
Gilbert FA (1927) On the occurrence of biflagellate swarm cells in certain
  myxomycetes.
\newblock {\em\JournalTitle{Mycologia}} 19:277--283.

\bibitem{Dieckmann2003}
Dieckmann CL (2003) Eyespot placement and assembly in the green alga
  \textit{Chlamydomonas}.
\newblock {\em\JournalTitle{Bioessays}} 25(4):410--416.

\bibitem{Wan2014b}
Wan KY, Goldstein RE (2014) Rhythmicity, recurrence, and recovery of flagellar
  beating.
\newblock {\em\JournalTitle{Physical Review Letters}} 113(23):238103.

\bibitem{Ruffer1987}
Ruffer U, Nultsch W (1987) Comparison of the beating of \textit{Cis}-flagella
  and \textit{Trans}-flagella of \textit{Chlamydomonas} cells held on
  micropipettes.
\newblock {\em\JournalTitle{Cell Motility and the Cytoskeleton}} 7(1):87--93.

\bibitem{Goldstein2009}
Goldstein RE, Polin M, Tuval I (2009) Noise and synchronization in pairs of
  beating eukaryotic flagella.
\newblock {\em\JournalTitle{Physical Review Letters}} 103(16):168103.

\bibitem{Bruot2012}
Bruot N, Kotar J, de~Lillo F, Lagomarsino MC, Cicuta P (2012) Driving potential
  and noise level determine the synchronization state of hydrodynamically
  coupled oscillators.
\newblock {\em\JournalTitle{Physical Review Letters}} 109(16):164103.

\bibitem{Geyer2013}
Geyer VF, J\"ulicher F, Howard J, Friedrich BM (2013) Cell-body rocking is a
  dominant mechanism for flagellar synchronization in a swimming alga.
\newblock {\em\JournalTitle{Proceedings of the National Academy of Sciences of
  the United States of America}} 110(45):18058--18063.

\bibitem{Quaranta2015}
Quaranta G, Aubin-Tam M, Tam D (2015) On the role of hydrodynamics vs
  intracellular coupling in synchronization of eukaryotic flagella.
\newblock {\em\JournalTitle{Phys. Rev. Lett.}} 115(23):238101.

\bibitem{Witman1993}
Witman GB (1993) \textit{Chlamydomonas} phototaxis.
\newblock {\em\JournalTitle{Trends Cell Biology}} 3(11):403--408.

\bibitem{Lewin1952}
Lewin RA (1952) Studies on the flagella of algae .1. general observations on
  \textit{Chlamydomonas moewusii} \text{Gerloff}.
\newblock {\em\JournalTitle{Biological Bulletin}} 103(1):74--79.

\bibitem{Hyams1978}
Hyams JS, Borisy GG (1978) Isolated flagellar apparatus of
  \textit{Chlamydomonas} - characterization of forward swimming and alteration
  of waveform and reversal of motion by calcium-ions invitro.
\newblock {\em\JournalTitle{Journal of Cell Science}} 33(OCT):235--253.

\bibitem{Wright1983}
Wright RL, Choknacki B, Jarvik JW (1983) Abnormal basal-body number, location,
  and orientation in a striated fiber-defective mutant of \textit{Chlamydomonas
  reinhardtii}.
\newblock {\em\JournalTitle{Journal of Cell Biology}} 96(6):1697--1707.

\bibitem{Wright1985}
Wright RL, Salisbury JL, Jarvik JW (1985) A nucleus-basal body connector in
  \textit{Chlamydomonas reinhardtii} that may function in basal body
  localization or segregation.
\newblock {\em\JournalTitle{Journal of Cell Biology}} 101(5):1903--1912.

\bibitem{Geimer2004}
Geimer S, Melkonian M (2004) The ultrastructure of the \textit{Chlamydomonas
  reinhardtii} basal apparatus: identification of an early marker of radial
  asymmetry inherent in the basal body.
\newblock {\em\JournalTitle{Journal of Cell Science}} 117(13):2663--2674.

\bibitem{Hoops1984}
Hoops HJ, Wright RL, Jarvik JW, Witman GB (1984) Flagellar waveform and
  rotational orientation in a \textit{Chlamydomonas} mutant lacking normal
  striated fibers.
\newblock {\em\JournalTitle{Journal of Cell Biology}} 98(3):818--824.

\bibitem{Niedermayer2008}
Niedermayer T, Eckhardt B, Lenz P (2008) Synchronization, phase locking, and
  metachronal wave formation in ciliary chains.
\newblock {\em\JournalTitle{Chaos}} 18(3):037128.

\bibitem{Elfring2009}
Elfring GJ, Lauga E (2009) Hydrodynamic phase locking of swimming
  microorganisms.
\newblock {\em\JournalTitle{Physical Review Letters}} 103(8):088101.

\bibitem{Stewart1978}
Stewart KD, Mattox KR (1978) Structural evolution in flagellated cells of
  green-algae and land plants.
\newblock {\em\JournalTitle{Biosystems}} 10(1-2):145--152.

\bibitem{Melkonian1980}
Melkonian M (1980) Ultrastructural aspects of basal body associated fibrous
  structures in green-algae - a critical-review.
\newblock {\em\JournalTitle{Biosystems}} 12(1-2):85--104.

\bibitem{Daugbjerg1992}
Daugbjerg N, Moestrup O (1992) Ultrastructure of \textit{Pyramimonas
  cyrtoptera} sp-nov \text{(Prasinophyceae)}, a species with $16$ flagella from
  nortrates\text{Foxe} basin, arctic \text{Canada}, including observations on
  growth rates.
\newblock {\em\JournalTitle{Canadian Journal of Botany-revue Canadienne De
  Botanique}} 70(6):1259--1273.

\bibitem{Conrad1939}
Conrad W (1939) Notes protistologique. xi. sur \textit{Pyramidomonas amylifera}
  n. sp.
\newblock {\em\JournalTitle{Bulletin R. Hist. Nat. Belg.}} 15:1--10.

\bibitem{Hargraves1980}
Hargraves P, Gardiner W (1980) The life history of \textit{Pyramimonas
  amylifera} \text{Conrad (Prasinophyceae)}.
\newblock {\em\JournalTitle{Journal of plankton research}} 2(2):99--108.

\bibitem{Hori1987}
Hori T, Moestrup O (1987) Ultrastructure of the flagellar apparatus in
  \textit{Pyramimonas octopus} \text{(Prasinophyceae)}.1. axoneme structure and
  numbering of peripheral doublets triplets.
\newblock {\em\JournalTitle{Protoplasma}} 138(2-3):137--148.

\bibitem{Sym1991a}
Sym SD, Piernaar RN (1991) Ultrastructure of \textit{Pyramimonas norrisii} sp
  nov \text{(Prasinophyceae)}.
\newblock {\em\JournalTitle{British Phycological Journal}} 26(1):51--66.

\bibitem{Manton1968}
Manton I (1968) Observations on the microanatomy of the type species of
  \textit{Pyramimonas} (\textit{P. tetrarhynchus} \text{Schmarda}).
\newblock {\em\JournalTitle{Proc. Linn. Soc. Lond}} 179(2):147--152.

\bibitem{Salisbury1978}
Salisbury JL, Floyd GL (1978) Calcium-induced contraction of rhizoplast of a
  quadriflagellate green-alga.
\newblock {\em\JournalTitle{Science}} 202(4371):975--977.

\bibitem{Collins1993}
Collins JJ, Stewart IN (1993) Coupled nonlinear oscillators and the symmetries
  of animal gaits.
\newblock {\em\JournalTitle{Journal of Nonlinear Science}} 3(3):349--392.

\bibitem{Schoner1990}
Schoner G, Jiang WY, Kelso JAS (1990) A synergetic theory of quadrupedal gaits
  and gait transitions.
\newblock {\em\JournalTitle{Journal of Theoretical Biology}} 142(3):359--391.

\bibitem{Okelly1983a}
O'Kelly CJ, Floyd GL (1983) Flagellar apparatus absolute orientations and the
  phylogeny of the green algae.
\newblock {\em\JournalTitle{Biosystems}} 16(3-4):227--251.

\bibitem{Moestrup1989}
Moestrup O, Hori T (1989) Ultrastructure of the flagellar apparatus in
  \textit{Pyramimonas octopus} \text{(Prasinophyceae)} .2. flagellar roots,
  connecting fibers, and numbering of individual flagella in green-algae.
\newblock {\em\JournalTitle{Protoplasma}} 148(1):41--56.

\bibitem{Norris1975}
Norris RE, Pearson BR (1975) Fine structure of \textit{Pyramimonas parkeae} new
  species \text{Chlorophyta Prasinophyceae}.
\newblock {\em\JournalTitle{Archiv fuer Protistenkunde}} 117(1-2):192--213.

\bibitem{Brown1976}
Brown DL, Massalski A, Paternaude R (1976) Organization of flagellar apparatus
  and associated cytoplasmic microtubules in quadriflagellate alga
  \textit{Polytomella agilis}.
\newblock {\em\JournalTitle{Journal of Cell Biology}} 69(1):106--125.

\bibitem{Salisbury1981}
Salisbury JL, Swanson JA, Floyd GL, Hall R, Maihle NJ (1981) Ultrastructure of
  the flagellar apparatus of the green alga \textit{Tetraselmis subcordiformis}
  - with special consideration given to the function of the rhizoplast and
  rhizanchora.
\newblock {\em\JournalTitle{Protoplasma}} 107(1-2):1--11.

\bibitem{Lembi1975}
Lembi CA (1975) Fine-structure of flagellar apparatus of \textit{Carteria}.
\newblock {\em\JournalTitle{Journal of Phycology}} 11(1):1--9.

\bibitem{Harris2009}
Harris EH (2009) {\em The \textit{Chlamydomonas} sourcebook. A comprehensive
  guide to biology and laboratory use.}

\bibitem{ProtistologyBook}
Hausman K, Radek R, eds. (2014) {\em Cilia and flagella, ciliates and
  Flagellates: ultrastructure and cell biology, function and systematics,
  symbiosis and biodiversity}.
\newblock (Schweizerbart Science Publishers).

\bibitem{Beech1991}
Beech PL, Heimann K, Melkonian M (1991) Development of the flagellar apparatus
  during the cell-cycle in unicellular algae.
\newblock {\em\JournalTitle{Protoplasma}} 164(1-3):23--37.

\bibitem{Sun2014}
Sun D, Roth S, Black MJ (2014) A quantitative analysis of current practices in
  optical flow estimation and the principles behind them.
\newblock {\em\JournalTitle{International Journal of Computer Vision}}
  106(2):115--137.

\bibitem{Daugbjerg1994}
Daugbjerg N, Moestrup O, Archtander P (1994) Phylogeny of the genus
  \textit{Pyramimonas} (\text{Prasinophyceae, Chlorophyta}) inferred from the
  rbc\text{L} gene.
\newblock {\em\JournalTitle{Journal of Phycology}} 30(6):991--999.

\bibitem{Okita2005}
Okita N, Isogai N, Hirono M, Kamiya R, Yoshimura K (2005) Phototactic activity
  in \textit{Chlamydomonas} 'non-phototactic' mutants deficient in
  \text{Ca}$^{2+}$-dependent control of flagellar dominance or in inner-arm
  dynein.
\newblock {\em\JournalTitle{Journal of Cell Science}} 118(3):529--537.

\bibitem{Kamiya1984}
Kamiya R, Witman GB (1984) Submicromolar levels of calcium control the balance
  of beating between the two flagella in demembranated models of
  \textit{Chlamydomonas}.
\newblock {\em\JournalTitle{Cell Motility and the Cytoskeleton}} 98(1):97--107.

\bibitem{Bessen1980}
Bessen M, Fay RB, Witman GB (1980) Calcium control of waveform in isolated
  flagellar axonemes of \textit{Chlamydomonas}.
\newblock {\em\JournalTitle{Journal of Cell Biology}} 86(2):446--455.

\bibitem{Mukundan2014}
Mukundan V, Sartori P, Geyer VF, J\"ulicher F, Howard J (2014) Motor regulation
  results in distal forces that bend partially disintegrated
  \textit{Chlamydomonas} axonemes into circular arcs.
\newblock {\em\JournalTitle{Biophysical Journal}} 106(11):2434--2442.

\bibitem{Melkonian1983a}
Melkonian M (1983) Functional and phylogenetic aspects of the basal apparatus
  in algal cells.
\newblock {\em\JournalTitle{Journal of Submicroscopic Cytology and Pathology}}
  15(1):121--125.

\bibitem{Salisbury1988}
Salisbury JL (1988) The lost neuromotor apparatus of \textit{Chlamydomonas} -
  rediscovered.
\newblock {\em\JournalTitle{Journal of Protozoology}} 35(4):574--577.

\bibitem{Gibbons1960}
Gibbons IR, Grimstone AV (1960) On flagellar structure in certain flagellates.
\newblock {\em\JournalTitle{Journal of Biophysical and Biochemical Cytology}}
  7(4):697--\&.

\bibitem{White1981}
White RB, Brown DL (1981) \text{ATP}ase activities associated with the
  flagellar basal apparatus of \textit{Polytomella}.
\newblock {\em\JournalTitle{Journal of Ultrastructure Research}}
  75(2):151--161.

\bibitem{Hayashi1998}
Hayashi M, Yagi T, Yoshimura K, Kamiya R (1998) Real-time observation of
  \text{Ca}$^{2+}$-induced basal body reorientation in \textit{Chlamydomonas}.
\newblock {\em\JournalTitle{Cell Motility and the Cytoskeleton}} 41(1):49--56.

\bibitem{Watson1975}
Watson MW (1975) Flagellar apparatus, eyespot and behavior of
  \textit{Microthamnion kuetzingianum} \text{(Chlorophyceae)} zoospores.
\newblock {\em\JournalTitle{Journal of Phycology}} 11(4):439--448.

\bibitem{Matsusaka1967}
Matsusaka T (1967) \text{ATP}ase activity in the ciliary rootlet of human
  retinal rods.
\newblock {\em\JournalTitle{The Journal of cell biology}} 33(1):203--8.

\bibitem{Cavaliersmith1982}
Cavalier-Smith T (1982) The origins of plastids.
\newblock {\em\JournalTitle{Biological Journal of the Linnean Society}}
  17(3):289--306.

\bibitem{Leliaert2012}
Leliaert F et~al. (2012) Phylogeny and molecular evolution of the green algae.
\newblock {\em\JournalTitle{Critical Reviews In Plant Sciences}} 31(1):1--46.

\bibitem{Buchheim1996}
Buchheim MA et~al. (1996) Phylogeny of the \text{Chlamydomonadales
  (Chlorophyceae)}: A comparison of ribosomal \text{RNA} gene sequences from
  the nucleus and the chloroplast.
\newblock {\em\JournalTitle{Molecular Phylogenetics and Evolution}}
  5(2):391--402.

\bibitem{McFadden1986}
Mcfadden GI, Hill D, Wetherbee R (1986) A study of the genus pyramimonas
  (\text{prasinophyceae}) from southeastern \text{Australia}.
\newblock {\em\JournalTitle{Nordic Journal of Botany}} 6(2):209--234.

\bibitem{Nozaki2003}
Nozaki H, Misumi O, Kuroiwa T (2003) Phylogeny of the quadriflagellate
  \text{Volvocales (Chlorophyceae)} based on chloroplast multigene sequences.
\newblock {\em\JournalTitle{Molecular Phylogenetics and Evolution}}
  29(1):58--66.

\bibitem{Smith2014}
Smith DR, Lee RW (2014) A plastid without a genome: Evidence from the
  nonphotosynthetic green algal genus \text{Polytomella}.
\newblock {\em\JournalTitle{Plant Physiology}} 164(4):1812--1819.

\bibitem{Brooks2014}
Brooks ER, Wallingford JB (2014) Multiciliated cells.
\newblock {\em\JournalTitle{Current Biology}} 24(19):R973--R982.

\bibitem{CosentinoLagomarsino2003}
Cosentino~Lagomarsino M, Jona P, Bassetti B (2003) Metachronal waves for
  deterministic switching two-state oscillators with hydrodynamic interaction.
\newblock {\em\JournalTitle{Physical Review E}} 68(2):021908.

\bibitem{Ma2014}
Ma R, Klindt GS, Riedel-Kruse IH, J\"ulicher F, Friedrich BM (2014) Active
  phase and amplitude fluctuations of flagellar beating.
\newblock {\em\JournalTitle{Physical Review Letters}} 113(4):048101.

\bibitem{Harz1991}
Harz H, Hegemann P (1991) Rhodopsin-regulated calcium currents in
  \textit{Chlamydomonas}.
\newblock {\em\JournalTitle{Nature}} 351(6326):489--491.

\bibitem{Jekely2015}
\text{J\'ekely} G, Paps J, Nielsen C (2015) The phylogenetic position of
  ctenophores and the origin(s) of nervous systems.
\newblock {\em\JournalTitle{Evodevo}} 6:1.

\bibitem{SCCAP_media}
(2015) http://www.sccap.dk/media/.

\bibitem{CCAP_media}
(2015) http://www.ccap.ac.uk/pdfrecipes.htm.

\end{thebibliography}

\end{article}

\end{document}